\documentclass[aps,showpacs,prb,twocolumn,superscriptaddress,floatfix,longbibliography]{revtex4-2}
\usepackage[pdftex]{graphicx}
\usepackage{amsbsy, amssymb, amsmath, bm, mathtools}
\usepackage{xspace}
\usepackage{bm}
\usepackage{color}
\usepackage{tabularx}
\usepackage{booktabs}
\usepackage[colorlinks=true,linkcolor=blue,citecolor=blue]{hyperref}
\usepackage{url}
\usepackage{threeparttable}
\usepackage{multirow}
\usepackage{enumerate}
\usepackage{enumitem}
\usepackage{color}
\usepackage{subfigure}  
\usepackage{epstopdf}
\usepackage{bbm}
\usepackage{graphicx}
\usepackage{hyperref}
\usepackage{threeparttable}
\usepackage{amsthm}
\usepackage{mathtools}
\usepackage{color}
\usepackage{tikz}
\usepackage{empheq}
\usepackage{adjustbox}
\usepackage{enumitem}
\usepackage{siunitx}
\usepackage{array}

\newcolumntype{L}{>{\raggedright\arraybackslash\hspace{6pt}}X<{\hspace{6pt}}}
\begin{document}
\title{Long-range self-avoiding walk in one dimension: a Monte Carlo study}

\author{Jiang Zhou}
\thanks{These two authors contributed equally to this paper.}
\affiliation{Department of Physics, Guizhou University, Guiyang 550025, China}
\author{Ziyu Liu}
\thanks{These two authors contributed equally to this paper.}
\affiliation{School of Emerging Technology, University of Science and Technology of China,
Hefei 230026, China}
\author{Pengcheng Hou}
\email{houpc@hfnl.cn}
\affiliation{Hefei National Laboratory, University of Science and Technology of China, Hefei 230088, China}
\author{ Zhijie Fan}
\email{zfanac@ustc.edu.cn}
\affiliation{Hefei National Laboratory, University of Science and Technology of China, Hefei 230088, China}
\affiliation{Hefei National Research Center for Physical Sciences at the Microscale and School of Physical Sciences, University of Science and Technology of China, Hefei 230026, China}
\affiliation{Shanghai Research Center for Quantum Science and CAS Center for Excellence in Quantum Information and Quantum Physics, University of Science and Technology of China, Shanghai 201315, China}
\author{Youjin Deng}
\email{yjdeng@ustc.edu.cn}
\affiliation{Department of Physics, Guizhou University, Guiyang 550025, China}
\affiliation{Hefei National Laboratory, University of Science and Technology of China, Hefei 230088, China}
\affiliation{Hefei National Research Center for Physical Sciences at the Microscale and School of Physical Sciences, University of Science and Technology of China, Hefei 230026, China}
\affiliation{Shanghai Research Center for Quantum Science and CAS Center for Excellence in Quantum Information and Quantum Physics, University of Science and Technology of China, Shanghai 201315, China}

\begin{abstract}
We study the one-dimensional long-range self-avoiding walk in the grand-canonical ensemble
where the statistical weight of a jump of length $r$ decays
algebraically as $r^{-(d+\sigma)}$.
Using large-scale Monte Carlo simulations with an efficient irreversible update scheme, 
we obtain high-precision estimates of the critical fugacity $z_c$, the universal Binder ratio $Q_N^c$, 
the correlation-length exponent $\nu$, and the anomalous dimension $\eta$.
For $\sigma>1$, the critical fugacity varies smoothly with $\sigma$, while
the Binder ratio and the critical exponents remain consistent with the short-range universality class.
For $\sigma \le 1$, by contrast, the results clearly depart from short-range behavior, 
identifying $\sigma=1$ as the boundary between long-range and short-range regimes.
In the long-range Wilson-Fisher regime with $1/2 < \sigma \leq 1$, 
$\eta$ agrees with the long-range Gaussian-fixed-point prediction $\eta_{\mathrm{GFP}}=2-\sigma$, 
whereas $Q_N^c$ and $\nu$ vary nontrivially with $\sigma$ and exhibit discontinuous jumps at $\sigma=1$.
These findings are in good agreement with the recently proposed  universality diagram in the $(d,\sigma)$ plane for long-range O$(n)$ models, with the self-avoiding walk
corresponding to the $n\to0$ limit.
\end{abstract}
\maketitle

\section{Introduction}
\label{sec:intro}

Long-range (LR) interactions with algebraic decay arise in a wide range of physical systems, from magnetic and dipolar media to trapped-ion, Rydberg-atom, and network systems~\cite{campa2014,defenu2023}. They provide a natural way to tune the effective geometric connectivity of a model and can change the nature of critical behavior relative to its short-range (SR) counterpart~\cite{dyson1969,thouless1969,maghrebi2017}. A canonical theoretical model is the LR-O$(n)$ model, in which the pair coupling decays as $J(r)\sim r^{-(d+\sigma)}$ in $d$ spatial dimensions.
For sufficiently small $\sigma$, the interaction decays slowly with distance $r$, and
the critical behavior is described by the LR Gaussian fixed point (GFP)
and the critical exponents take the mean-field (MF) values.
For sufficiently large $\sigma$, the critical behavior of the SR universality is recovered.
In the nonclassical SR regime, critical behavior is described by the Wilson--Fisher (WF) fixed point~\cite{WilsonFisher1972} for SR interactions.
In between, a LR-WF fixed point is expected, for which the critical exponents
can depend on $\sigma$, $d$, and $n$.
Determining the boundaries between these regimes
and the behavior at the boundaries themselves has therefore become a basic question in the theory of LR critical phenomena.

\begin{figure}[t]
    \centering
    \includegraphics[width=0.9\linewidth]{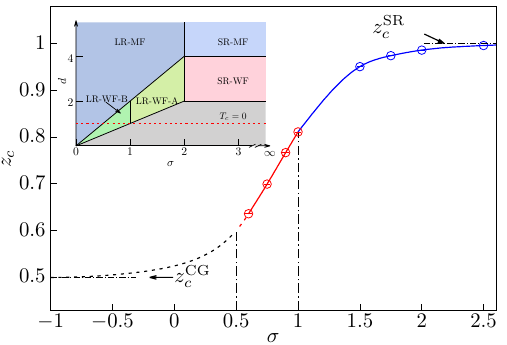}
    \caption{Critical fugacity $z_c$ as a function of $\sigma$ for the 1D LR-SAW.
    The inset shows the recently proposed diagram of universality classes for
    the LR-O$(n)$ model
    in the $(\sigma,d)$ plane, adapted from Ref.~\cite{xiao2025universalitydiagramphasetransitions}.
    The diagram
    consists of six regimes: short-range mean-field (SR-MF),
    long-range mean-field (LR-MF), short-range Wilson--Fisher (SR-WF),
    long-range Wilson--Fisher-A (LR-WF-A), long-range Wilson--Fisher-B (LR-WF-B), and zero-temperature ($T_c=0$) regimes. The red dashed line in the inset marks $d=1$,
    corresponding to the 1D long-range self-avoiding walk (LR-SAW) studied in the main panel.}
    \label{fig:phasediag}
\end{figure}

The location of the LR--SR boundary has been debated since the early
renormalization-group (RG) analysis of Fisher \textit{et al.},
which predicted $\sigma_*=2$ using an $\epsilon$-expansion~\cite{fisher1972}.
Sak subsequently proposed $\sigma_*=2-\eta_{\mathrm{SR}}$~\cite{sak1973}.
Here, $\eta$ is the anomalous dimension governing the critical
two-point correlation function, $g(r)\sim r^{2-d-\eta}$,
and $\eta_{\mathrm{SR}}$ denotes its value in the SR universality class.
The LR--SR crossover has also been investigated numerically in two-dimensional (2D) LR-Ising models~\cite{luijten2002,PhysRevE.89.062120}
and through functional RG studies of LR-O$(n)$ models~\cite{Defenu2015}.
Related field-theoretic work has examined the LR-Ising crossover~\cite{Behan2017}.
Recent large-scale numerical studies~\cite{xiao2025sakscriterionstatisticalmodels,xiao_two-dimensional_2024,yao2025spontaneoussymmetrybreakingtwodimensional,liu2025twodimensionalpercolationmodellongrange,liu2026}
and perturbative RG calculations~\cite{li20264epsilonexpansionlongrangeinteracting,li2026perturbativerenormalizationuniversalitydiagram,li2026a} have renewed this discussion and motivated the recently proposed universality diagram in the $(d,\sigma)$ plane, as shown in the inset of Fig.~\ref{fig:phasediag}.
The detailed classification and its theoretical basis are discussed in Ref.~\cite{xiao2025universalitydiagramphasetransitions}.
In particular, the $d=2$ line in this diagram has been tested in high-precision simulations of LR-O$(n)$ models and LR-percolation~\cite{xiao_two-dimensional_2024,yao2025spontaneoussymmetrybreakingtwodimensional,liu2025twodimensionalpercolationmodellongrange,xiao2025sakscriterionstatisticalmodels}.
Along the $d=1$ line, marked by the red dashed line in the inset,
the proposed diagram identifies a zero-temperature ($T_c=0$)
regime for $\sigma>1$, where the system has no nontrivial phase transition
at finite temperature ($T>0$).
The interval $1/2<\sigma<1$ is termed the LR-WF-B regime. In this interval,
the critical behavior is governed by an LR-WF fixed point,
at which the correlation-length exponent $\nu$ varies nontrivially with $\sigma$,
whereas the anomalous dimension remains at $\eta=\eta_{\mathrm{GFP}}$, with $\eta_{\mathrm{GFP}}=2-\sigma$ denoting the LR-GFP prediction.
In the LR-MF regime for $0<\sigma<1/2$, both critical exponents
take their GFP values, $\nu=\nu_{\mathrm{GFP}}=1/\sigma$ and $\eta=\eta_{\mathrm{GFP}}$.
At the boundary $\sigma=1$, the nature of the phase transition depends on
the number $n$ of spin components,
and, at $\sigma=1/2$, logarithmic corrections can arise.


The one-dimensional (1D) LR-Ising model with ferromagnetic
couplings $J(r)\sim r^{-(1+\sigma)}$, corresponding to the
$n=1$ case of the LR-O$(n)$ model, provides a well-studied
benchmark for LR critical phenomena. 
LR couplings in the 1D LR-Ising model can suppress fluctuations,
allowing spontaneous symmetry breaking.
Dyson first rigorously established the existence of a finite-temperature
phase transition, with spontaneous magnetization at low
temperatures for $0<\sigma<1$~\cite{dyson1969}. For $\sigma>1$,
spontaneous magnetization is absent at any finite
temperature~\cite{dyson1969nonexistence,ruelle1968}.
At the boundary $\sigma=1$, the model undergoes a
finite-temperature phase transition with Berezinskii-Kosterlitz-Thouless-like (BKT-like) features,
including an essential singularity in the free energy~\cite{frohlich1982,PhysRevLett.37.1577}.
However, unlike the standard BKT transition in the 2D
XY model~\cite{KosterlitzThouless1973}, this transition is accompanied by a discontinuous
jump in the spontaneous magnetization~\cite{thouless1969,aizenman1988}.
Large-scale Monte Carlo simulations have confirmed that, in the regime $0<\sigma\leq1/2$, the system lies in the classical (mean-field) universality class, with the anomalous dimension $\eta$ fixed at the LR-GFP value $\eta_{\mathrm{GFP}}=2-\sigma$~\cite{Luijten1997}.
In the nonclassical interval $1/2<\sigma<1$, recent functional
renormalization-group calculations for the 1D LR-Ising model
within the local potential approximation (LPA) yield estimates
of the correlation-length exponent $\nu$ that closely agree
with those obtained from a real-space RG analysis of Dyson's
hierarchical model~\cite{dyson1969,Pagni2025}.
These calculations recover $1/\nu\to1/2$ as $\sigma\to1/2^+$,
consistent with the LR-GFP relation $1/\nu_{\mathrm{GFP}}=\sigma$,
and yield $1/\nu\to0$ as $\sigma\to1^-$, consistent with
the approach to BKT-like criticality at $\sigma=1$.
Recent conformal field theory (CFT) work has proposed a dual description of the
critical 1D LR-Ising model that becomes weakly coupled near
$\sigma=1$, allowing perturbative calculations of CFT
data~\cite{Benedetti2025}.

The 1D LR self-avoiding walk (LR-SAW), corresponding to the
$n\to0$ limit of the LR-O$(n)$ model~\cite{deGennes1972,de1979scaling}, provides a complementary
test of this universality picture.
We use the standard grand-canonical convention in which an $N$-step walk carries a statistical weight $z^N$, with $z$ the fugacity per step (or bond)~\cite{Fang_2021}. The critical fugacity $z_c$ is the boundary of convergence of the corresponding walk generating function.
For the nearest-neighbor (NN) SAW on a finite ring of length $L$---i.e.,
the SR $\sigma\to\infty$ limit,
self-avoidance permits only the two monotone walks of each length, giving $\mathcal Z_L^{\mathrm{SR}}(z)=1+2\sum_{N=1}^{L-1}z^N$ in 1D. In the thermodynamic limit, the corresponding series converges
for $z<1$ and becomes singular at $z=1$, yielding
$z_c^{\mathrm{SR}}=1$. Throughout this paper, descriptive labels such as SR are written as subscripts
unless a quantity already carries an intrinsic subscript, in which case
they are placed as superscripts, as in $z_c^{\mathrm{SR}}$.
In the high-temperature graph expansion of the NN Ising [O$(1)$] model, the corresponding bond activity is $z=\tanh\beta$.
Thus, the limit $z_c\to1$ corresponds to $T_c=1/\beta_c\to0$,
in the $T_c=0$ regime of the proposed universality diagram.
This naturally raises the question of how the introduction of algebraically distributed jumps changes the critical fugacity $z_c$,
the critical exponents, and other universal ratios.

Let $R_N$ denote a typical end-to-end distance of an $N$-step SAW.
Its asymptotic growth, $R_N\sim N^\nu$, defines the correlation-length exponent $\nu$.
For SR-SAW, Flory's mean-field treatment, based on an energy-entropy variation argument, 
gives $\nu_{\mathrm{F}}=3/(d+2)$ for $d\leq d_c=4$~\cite{Flory1953,de1979scaling}.
Despite its simplicity, this mean-field approximation is surprisingly good:
it gives the exact values $\nu=1$ for $d=1$, $\nu=3/4$ for $d=2$~\cite{Nienhuis1982}, and $\nu=1/2$ for
$d\geq4$; for $d=3$, $\nu_{\mathrm{F}}=3/5$ is remarkably close to the best numerical value
$\nu=0.587\,597\,0(4)$~\cite{Clisby2010,Nathan2016}.
A Flory-type extension to the LR-SAW predicts
$\nu_{\mathrm{F}}=3/(d+\sigma)$ for $d/2<\sigma\leq2$,
whereas the SR expression $\nu_{\mathrm{F}}=3/(d+2)$ is recovered
for $\sigma>2$. This approximation therefore predicts an LR--SR
crossover at $\sigma_*=2$
~\cite{de1979scaling,halley1985node,grassberger1985critical}.
Early Monte Carlo studies reported good agreement between their
estimates of the critical exponents and the corresponding
Flory-type predictions~\cite{halley1985node,grassberger1985critical}.
Nevertheless, more recently, large-scale fixed-$N$ simulations of the 1D LR-SAW
provided further numerical evidence for the crossover at
$\sigma_*=1$ and revealed strong corrections to scaling near the
boundary~\cite{sarkar2025longrangeshortrange}.
This finding is consistent with the universality-class diagram in the inset of Fig.~\ref{fig:phasediag}
and also agree with Sak's criterion $\sigma_*=2-\eta_{\mathrm{SR}}$~\cite{sak1973}.
Let $G_L(r;z)$ denote the unnormalized end-to-end correlation function at distance $r$.
For the NN-SAW at $z_c^{\mathrm{SR}}=1$,
$G_L(r;1)\asymp 1$; after normalization over all endpoint
positions, the corresponding endpoint probability distribution
forms an $L$-dependent plateau of order $L^{-1}$.
Comparing $G_L(r;1)\asymp 1$ with the critical algebraic form
$G(r;z_c)\sim r^{2-d-\eta}$ gives $\eta_{\mathrm{SR}}=1$ in $d=1$,
and Sak's criterion therefore yields
$\sigma_*=2-\eta_{\mathrm{SR}}=1$.

\begin{figure}
    \centering
    \includegraphics[width=0.95\linewidth]{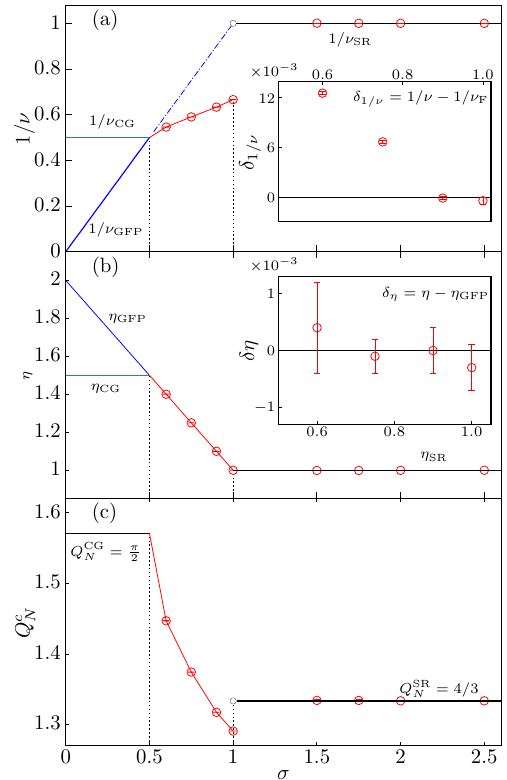}
    \caption{The inverse correlation-length exponent $1/\nu$, anomalous dimension $\eta$, and critical Binder ratio $Q_N^c$ as functions of $\sigma$ are shown in panels (a), (b), and (c), respectively. In panel (a), $1/\nu$ agrees with the SR value $1/\nu_{\mathrm{SR}}=1$ for $\sigma>1$ (black solid line). In the nonclassical LR regime $1/2<\sigma\leq1$, $1/\nu$ varies nontrivially with $\sigma$: it is distinct from both the SR value and the LR-GFP prediction
    $1/\nu_{\mathrm{GFP}}=\sigma$. It also deviates from Flory's prediction $1/\nu_{\mathrm{F}}=(1+\sigma)/3$, although the deviation cannot be quantitatively resolved
    for $\sigma=0.9$ and $1$, as shown in the inset of panel (a). For $0<\sigma<1/2$, the LR Gaussian fixed point predicts $1/\nu_{\mathrm{GFP}}=\sigma$ (blue solid line), and the complete-graph asymptotics
    give $1/\nu_{\mathrm{CG}}=d/2=1/2$ (green solid line).
    In panel (b), $\eta$ takes the SR value $\eta_{\mathrm{SR}}=1$ for $\sigma>1$. In the nonclassical LR regime $1/2<\sigma\leq1$, the estimates of $\eta$ agree well with the LR-GFP prediction $\eta_{\mathrm{GFP}}=2-\sigma$ (red solid line). The inset shows the corresponding deviation $\delta\eta=\eta-\eta_{\mathrm{GFP}}$, which remains within the statistical uncertainties.
    In the LR mean-field regime $0<\sigma<1/2$, $\eta$ takes the GFP value $\eta_{\mathrm{GFP}}$ (blue solid line). The green solid line marks the complete-graph finite-size scaling value $\eta_{\mathrm{CG}}=2-1/\nu_{\mathrm{CG}}=3/2$. In panel (c), $Q_N^c$ is consistent with the SR universal value $Q_N^{\mathrm{SR}}=4/3$ for $\sigma>1$, takes nontrivial values for $1/2<\sigma\leq1$,
    and reduces to the complete-graph value $Q_N^{\mathrm{CG}}=\pi/2$ for $\sigma<1/2$.
    At the boundary $\sigma=1$, the LR-SAW exhibits a nontrivial second-order phase transition,
    and both $Q_N^c$ and $1/\nu$ exhibit discontinuous jumps from their SR values.}
    \label{fig:nu_eta_QN}
\end{figure}

\begin{table}[t]
\caption{Summary of the critical fugacity $z_c$, the universal critical
Binder ratio $Q_N^c$, the anomalous dimension $\eta$, and
the inverse correlation-length exponent $1/\nu$.}\label{table:summary}
\begin{tabular*}{\columnwidth}{@{\extracolsep{\fill}}lllll}
\hline\hline
$\sigma$ & $z_c$ &  $Q_N^c$ &  $\eta$     &    $1/\nu$ \\ \hline
   $\mathrm{NN}$ & 1 & 4/3 & 1 & 1 \\
    2.5 & 0.995\,593\,5(2) & 1.333\,39(5) & 1.000(1) & 0.999\,9(3) \\
    2.0 & 0.985\,821\,3(3) & 1.333\,5(2) & 0.999\,6(5) & 1.000\,1(3)  \\
    1.75 & 0.973\,895\,0(5) & 1.334(1) & 0.999(1) & 1.000\,1(4) \\
    1.5 & 0.950\,482\,47(8) & 1.334(1) & 0.999(1) & 1.000\,6(8) \\
    \hline
    1.0 & 0.810\,889\,3(4) & 1.290\,57(2)   & 0.999\,7(4)  & 0.666\,2(4) \\
    0.9 & 0.766\,704\,8(4) & 1.317\,20(3)   & 1.100\,0(4)  & 0.633\,2(2) \\
    0.75& 0.699\,573\,4(7) & 1.374\,2(2)    & 1.249\,9(3)  & 0.590\,0(2) \\
    0.6 & 0.636\,134(2)   & 1.446\,9(2)    & 1.400\,4(8)  & 0.545\,9(2) \\
\hline\hline
\end{tabular*}
\end{table}

In this work, we perform a Monte Carlo study of the 1D LR-SAW in the grand-canonical ensemble using an efficient irreversible algorithm.
Large-scale simulations are performed with periodic boundary conditions for linear sizes up to $L=2^{22}$.
We determine the critical fugacity $z_c$, the critical Binder ratio $Q_N^c$, the correlation-length exponent $\nu$, and the anomalous dimension $\eta$ over a range of decay exponents $\sigma>1/2$,
since the critical behavior for $\sigma\leq1/2$ is mean-field-like.
The main results are shown in Figs.~\ref{fig:phasediag} and~\ref{fig:nu_eta_QN}, as well as in Table~\ref{table:summary}.
The critical fugacity $z_c$ is a smooth function of $\sigma$ (Fig.~\ref{fig:phasediag}).
For comparison, setting $\sigma=-1$ in the bare power-law
weight gives uniform couplings on a finite set of sites,
providing a complete-graph reference.
With each bond assigned weight $2z/(L-1)$, this reference
model has critical fugacity $z_c^{\mathrm{CG}}=1/2$ in the
thermodynamic limit~\cite{slade2020completegraph,Deng_2019}.
The normalized and periodized LR-SAW model studied here
is defined only for $\sigma>0$.
Except in the NN limit, where $z_c(\sigma\to\infty)=1$,
the critical fugacity satisfies $z_c<1$ (Table~\ref{table:summary}). 
Nevertheless, Fig.~\ref{fig:nu_eta_QN} and Table~\ref{table:summary} show
that, for $\sigma>1$,
the Binder ratio $Q_N^c$, the inverse correlation-length exponent $1/\nu$, and the anomalous dimension $\eta$ are all consistent with the exact NN values.
For $1/2<\sigma\leq1$, $\eta$ agrees well with the LR-GFP value $\eta_{\mathrm{GFP}}=2-\sigma$,
with uncertainties in the fourth decimal place.
However, $Q_N^c$ and $1/\nu$ are nontrivial functions of $\sigma$ and, at the boundary $\sigma_*=1$,
display significant jumps from $(Q_N^c,1/\nu)=(1.290\,57(2),0.666\,2(4))$ at
$\sigma=1$ to $(4/3,1)$ for $\sigma=1^+$.
For $1/2<\sigma\leq1$, the estimated values of $1/\nu$ are again surprisingly close to
the Flory-type mean-field approximation $1/\nu_{\mathrm{F}}=(d+\sigma)/3$.
As shown in Table~\ref{table:summary} and the inset of Fig.~\ref{fig:nu_eta_QN}(a),
the deviation $1/\nu-1/\nu_{\mathrm{F}}$ can be resolved for $\sigma=0.75$ and $0.6$,
while it is  within the statistical error bar for $\sigma=0.9$ and $1$.
For $1<\sigma<2$, the numerical results remain at the SR values
and  Flory mean-field prediction is no longer applicable.
In short, the regime $\sigma>1$ belongs to the SR universality class,
whereas $1/2<\sigma\leq1$ displays nonclassical LR critical behavior,
in line with the recently proposed universality diagram in the inset of Fig.~\ref{fig:phasediag}.
At the boundary $\sigma_*=1$, unlike the LR-Ising model, which has a BKT-like transition,
the LR-SAW exhibits a second-order phase transition with $1/\nu=0.666\,2(4)\approx2/3$
and $\eta=0.999\,7(4)$; no clear evidence of logarithmic
corrections is found over the accessible system sizes.

The remainder of this paper is organized as follows. In Sec.~\ref{sec:model}, we define the 1D LR-SAW model and specify the normalization convention for the long-range jump kernel. In Secs.~\ref{sec:algorithms} and~\ref{sec:observables}, we describe the Monte Carlo algorithms and introduce the physical observables and finite-size scaling forms used to determine the critical fugacity and critical exponents. In Sec.~\ref{sec:findings}, we present the main numerical results. Finally, Sec.~\ref{sec:conclusion} summarizes the conclusions.

\section{Model, Algorithm, and Observables}

\subsection{Model}
\label{sec:model}

We consider a 1D LR-SAW on a periodic ring of
$L$ sites, labelled by $0,1,\ldots,L-1$. An $N$-step SAW is represented
by
\begin{equation}
    \omega_N=(x_0,x_1,\ldots,x_N).
\end{equation}
Here $x_i\in\{0,1,\ldots,L-1\}$ is the position after $i$ steps, and
the walk is rooted at $x_0=0$. Self-avoidance requires $x_i\neq x_j$
for $i\neq j$, and therefore $0\leq N\leq L-1$.
Under periodic boundary conditions, each step is specified
by a signed pre-modulo displacement $\pm\ell_i$, with $\ell_i\geq1$, so that
$x_{i+1}\equiv x_i\pm\ell_i\pmod L$.
Only its two endpoints are counted as visited.

The discrete step-length distribution is obtained by
discretizing the continuous power-law density
\begin{equation}
    \rho_\sigma(r)=C_\sigma r^{-(1+\sigma)},
    \qquad r\geq\frac12 ,
\end{equation}
where $C_\sigma$ is a $\sigma$-dependent normalization constant.
The normalization over the two possible directions,
\begin{equation}
    2\int_{1/2}^{\infty}\rho_\sigma(r)\,dr=1,
\end{equation}
fixes
\begin{equation}
    C_\sigma=\frac{\sigma}{2^{\sigma+1}}.
\end{equation}
The probability of an integer step length $\ell_i$ is the
$\rho_\sigma$ weight of the corresponding unit interval:
\begin{equation}
\label{eq:step_kernel}
\begin{split}
    P(\ell_i)
    ={}&2\int_{\ell_i-\frac12}^{\ell_i+\frac12}
    \rho_\sigma(r)\,dr
    \\
    ={}&(2\ell_i-1)^{-\sigma}
    -(2\ell_i+1)^{-\sigma}.
\end{split}
\end{equation}
Thus $P(\ell_i)$ is the normalized step-length probability,
while a specified directed step has probability $P(\ell_i)/2$.
In the grand-canonical ensemble, the number
of steps $N$ is allowed to fluctuate, and a walk with $N$ steps acquires
a fugacity factor $z^N$~\cite{Fang_2021}. Accordingly, a step of length
$\ell_i$ carries the statistical weight $zP(\ell_i)$.
The statistical weight of an $N$-step walk is therefore
\begin{equation}
    \pi_N(\omega_N)
    =z^N\prod_{i=0}^{N-1}P(\ell_i).
\end{equation}
The grand-canonical partition function is then
\begin{equation}
\label{eq:partition_function}
    \mathcal Z_L(z)
    =1+\sum_{N=1}^{L-1}
    \sum_{\omega_N}
    \pi_N(\omega_N),
\end{equation}
where the inner sum counts all legal signed pre-modulo
displacement sequences and the first term represents
the empty walk.
In the limit $\sigma\to\infty$, $P(1)\to1$, while
$P(\ell)\to0$ for every $\ell>1$.
Thus, for $L>2$, the model reduces to the
ordinary 1D NN-SAW,
\begin{equation}
    \mathcal Z_L^{\rm SR}(z)=1+2\sum_{N=1}^{L-1}z^N.
\end{equation}
For finite $L$, $\mathcal Z_L^{\rm SR}(z)$ is a polynomial and has no
singularity. In the thermodynamic limit, its radius of convergence is
$z_c^{\mathrm{SR}}=1$.

\subsection{Algorithm}
\label{sec:algorithms}

We simulate the 1D LR-SAW in the grand-canonical ensemble
using an irreversible algorithm~\cite{hu2017irreversible} developed from
the Berretti--Sokal (BS) update~\cite{berretti1985new}.
For an addition proposal,
we draw a uniform random number $0<u\leq1$ and set~\cite{Grassberger_2013}
\begin{equation}
    r=\frac{1}{2}u^{-1/\sigma}.
\end{equation}
The sampled value $r$ is rounded to the nearest positive
integer $\ell$, reproducing the step-length distribution $P(\ell)$ in
Eq.~\eqref{eq:step_kernel}. Either sign is then chosen with equal
probability, so that the directed step $\pm\ell$ has probability
$P(\ell)/2$ and its endpoint is reduced modulo $L$.

In the reversible BS algorithm~\cite{berretti1985new}, addition and deletion are selected with
equal probability. Addition proposes a directed step with probability
$P(\ell)/2$, whereas deletion removes the last step. For a legal addition,
the ratio of the target weights is
$\pi_{N+1}/\pi_N=zP(\ell)$. The factor $P(\ell)$ therefore cancels between
the target-weight ratio and the proposal probability, yielding the
acceptance probabilities
\begin{equation}
\label{eq:LRBSacc}
    A_+=\min\{1,2z\},
    \qquad
    A_-=\min\left\{1,\frac{1}{2z}\right\},
\end{equation}
for addition and deletion, respectively.

The irreversible algorithm~\cite{hu2017irreversible} retains these acceptance probabilities, but
the current mode, rather than a new random choice, determines which action
is attempted: the walk grows only in mode $(+)$ and shrinks only in mode
$(-)$. One update proceeds as follows:
\begin{itemize}
\item In mode $(+)$, propose an addition by the sampling procedure above.
If the new endpoint is unoccupied, accept the addition with probability
$A_+$ and remain in mode $(+)$. If the endpoint is occupied or the
addition is rejected, switch to mode $(-)$.
\item In mode $(-)$, delete the last step with probability $A_-$ and
remain in mode $(-)$ after an accepted deletion. A rejected deletion
switches the walk to mode $(+)$. The empty walk also switches directly
to mode $(+)$.
\end{itemize}

For a legal addition of a directed step of length $\ell$, the proposal
and acceptance probabilities in mode $(+)$ are $P(\ell)/2$ and $A_+$,
respectively. The reverse deletion in mode $(-)$ requires no step proposal
and is accepted with probability $A_-$. The corresponding currents obey
\begin{equation}
    \frac{\pi_N}{2}\,\frac{P(\ell)}{2}\,A_+
    =
    \frac{\pi_{N+1}}{2}\,A_-,
\end{equation}
Here $\pi_{N+1}/\pi_N=zP(\ell)$, while Eq.~\eqref{eq:LRBSacc} implies
$A_+=2zA_-$. The resulting Markov chain does not obey detailed balance but
maintains global balance through the probability currents associated with
accepted updates and mode switches.
By suppressing diffusive backtracking in the walk length, this irreversible
construction can substantially improve the sampling efficiency over the
reversible BS algorithm~\cite{hu2017irreversible}.

\subsection{Observables}
\label{sec:observables}

For each sampled walk $\omega$, we directly measure the walk length,
or number of steps,
\begin{equation}
    \mathcal N=|\omega|,
\end{equation}
and the empty-walk indicator,
\begin{equation}
    \mathcal D_0=\mathbf 1_{\mathcal N=0}.
\end{equation}
From these measured quantities, we sample the following observables:

(a) Binder ratio.
To locate the critical fugacity, we use the dimensionless Binder ratio
\begin{equation}
\label{eq:QNdef}
    Q_N=\frac{\langle\mathcal N^2\rangle}
    {\langle\mathcal N\rangle^2}.
\end{equation}
Here and below, $\langle\cdot\rangle$ denotes a grand-canonical
ensemble average.
At criticality, $Q_N$ is expected to approach the universal limit
\begin{equation}
    Q_N^c\equiv\lim_{L\to\infty}Q_N(z_c,L).
\end{equation}
Consequently, crossings of $Q_N$ for different system sizes provide
an estimator of the critical fugacity. In the 1D SR-SAW model at
$z_c=1$, one has
\begin{equation}
\label{eq:QNSR}
\begin{aligned}
Q_N^{\mathrm{SR}}
&=\lim_{L\to\infty}
\frac{\langle\mathcal N^2\rangle}{\langle\mathcal N\rangle^2} \\
&=\lim_{L\to\infty}
\frac{L(L-1)/3}{\left[L(L-1)/(2L-1)\right]^2} \\
&=\lim_{L\to\infty}
\frac{(2L-1)^2}{3L(L-1)}
=\frac{4}{3}.
\end{aligned}
\end{equation}
Comparison with this exact value provides a diagnostic of whether the
model belongs to the SR universality class.

(b) Mean walk length.
We define
\begin{equation}
    N\equiv\langle\mathcal N\rangle.
\end{equation}
This energy-like observable is used to estimate the thermal scaling
exponent $y_t=1/\nu$. Here $N$ denotes the ensemble mean, whereas the
subscript in $\omega_N$ labels the fixed length of an individual walk.
At criticality, it scales as
\begin{equation}
    N\sim L^{y_t}\sim L^{1/\nu}.
\end{equation}
For the 1D SR-SAW, $N_{\mathrm{SR}}\sim L$, giving
$\nu_{\mathrm{SR}}=1$.

(c) Empty-walk probability.
We define
\begin{equation}
    D_0\equiv\langle\mathcal D_0\rangle.
\end{equation}
This observable is used to estimate the anomalous dimension. Since the
empty walk has unit statistical weight, $D_0=\mathcal Z_L^{-1}$.
Equivalently, the
susceptibility-like observable
\begin{equation}
\label{eq:chidef}
    \chi\equiv\mathcal Z_L=\frac{1}{D_0}
\end{equation}
has the finite-size scaling form
\begin{equation}
    \chi\sim L^{2-\eta}=L^{2y_h-1},
    \qquad
    D_0\sim L^{\eta-2},
\end{equation}
where the second equality in the scaling exponent of $\chi$ follows
from $2-\eta=2y_h-d$ with $d=1$. For the 1D SR-SAW,
$\eta_{\mathrm{SR}}=1$ and $y_h^{\mathrm{SR}}=1$, so that
$\chi_{\mathrm{SR}}\sim L$ and $D_0^{\mathrm{SR}}\sim L^{-1}$.
These exact SR results are derived in Appendix~\ref{app:sr_exact}.

\section{Numerical results}
\subsection{Main results}
\label{sec:findings}
This subsection summarizes the central results of this work. The numerical estimates of the critical fugacity $z_c$, the universal critical Binder ratio $Q_N^c$, and the critical exponents $1/\nu$ and $\eta$ are listed in Table~\ref{table:summary}. The resulting critical line and the corresponding universality class assignment are summarized in Figs.~\ref{fig:phasediag} and~\ref{fig:nu_eta_QN}.

For $\sigma>1$, the LR tail shifts the nonuniversal critical fugacity away from the NN value: $z_c$ decreases as $\sigma$ approaches $1$ and remains below $z_c^{\mathrm{SR}}=1$ (Fig.~\ref{fig:phasediag}). In contrast, $Q_N^c$, $1/\nu$, and $\eta$ remain consistent with their SR values, $Q_N^{\mathrm{SR}}=4/3$, $1/\nu_{\mathrm{SR}}=1$, and $\eta_{\mathrm{SR}}=1$ (Fig.~\ref{fig:nu_eta_QN}). Thus, the nonlocal step distribution produces a nontrivial critical fugacity without changing the universal critical behavior: throughout the simulated $\sigma>1$ range, the model remains in the SR universality class and in the $T_c=0$ regime of the 1D LR-O$(n)$ universality diagram.

For $1/2<\sigma\leq1$, the numerical results change qualitatively. At $\sigma=1$, $Q_N^c$ and $1/\nu$ exhibit discontinuous jumps from their SR values.
Upon decreasing $\sigma$ below $1$, $Q_N^c$ increases monotonically, whereas $1/\nu$ decreases monotonically [panels (a) and (c) of Fig.~\ref{fig:nu_eta_QN}]. The estimates of $1/\nu$ are clearly distinct from the LR-GFP prediction $1/\nu_{\mathrm{GFP}}=\sigma$. As $\sigma$ approaches $1$ from below, the numerical estimates of $1/\nu$ remain close to the Flory-type expression $1/\nu_{\mathrm{F}}=(1+\sigma)/3$. In particular, for $\sigma=0.9$ and $1.0$, 
the discrepancy cannot be resolved within the quoted error bars [inset of Fig.~\ref{fig:nu_eta_QN}(a)]. This comparison is stated only as a numerical observation.
By contrast, $\eta$ is consistent with the LR-GFP prediction $\eta_{\mathrm{GFP}}=2-\sigma$ for all simulated values of $\sigma$ in $1/2<\sigma\le1$ [inset of Fig.~\ref{fig:nu_eta_QN}(b)] and connects continuously to the SR value $\eta_{\mathrm{SR}}=1$ at $\sigma=1$.
The combined behavior of $Q_N^c$, $1/\nu$, and $\eta$ identifies $\sigma=1$ as the LR--SR boundary and places $1/2<\sigma<1$ in the LR-WF-B regime~\cite{xiao2025universalitydiagramphasetransitions}.

For completeness, Fig.~\ref{fig:nu_eta_QN} also shows the theoretical
predictions for the LR mean-field (LR-MF) regime $0<\sigma<1/2$.
In this regime, critical phenomena, particularly finite-size scaling behaviors,
contain coexisting LR-GFP and complete-graph (CG) contributions. The singular
free-energy density is correspondingly described by a two-scale form
with separate GFP and CG scaling functions
~\cite{Fang_2021,Fang2022,Fang2023,
xiao2025universalitydiagramphasetransitions,liu2026}.
Denoting the thermal and magnetic scaling fields by $t$ and $h$,
respectively, this form can be written as
\begin{equation}
\begin{split}
f_{\mathrm{s}}(t,h,L)={}&L^{-d}\widetilde f_{\mathrm{GFP}}
\left(tL^{y_t^{\mathrm{GFP}}},hL^{y_h^{\mathrm{GFP}}}\right) \\
&+L^{-d}\widetilde f_{\mathrm{CG}}
\left(tL^{y_t^{\mathrm{CG}}},hL^{y_h^{\mathrm{CG}}}\right).
\end{split}
\end{equation}
For $X\in\{\mathrm{GFP},\mathrm{CG}\}$, the renormalization and
critical exponents are related by $y_t^X=1/\nu_X$ and
$y_h^X=(d+2-\eta_X)/2$. For the LR-SAW,
\begin{equation}
\begin{aligned}
\left(y_t^{\mathrm{GFP}},y_h^{\mathrm{GFP}}\right)
&=\left(\sigma,\; \frac{d+\sigma}{2}\right), \\
\left(y_t^{\mathrm{CG}},\;\; y_h^{\mathrm{CG}} \; \right)
&=\left(\frac{d}{2},\;\;\; \frac{3d}{4} \;\; \right).
\end{aligned}
\end{equation}
Equivalently,
\begin{equation}
\begin{aligned}
\left(\nu_{\mathrm{GFP}},\eta_{\mathrm{GFP}}\right)
&=\left(\frac{1}{\sigma}, \; 2-\sigma\right), \\
\left(\nu_{\mathrm{CG}},\;\; \eta_{\mathrm{CG}} \;\; \right)
&=\left(\frac{2}{d},\; 2-\frac{d}{2}\right).
\end{aligned}
\end{equation}
The two exponent sets coincide at $\sigma=d/2$.
For $d=1$, the CG reference values shown in
Fig.~\ref{fig:nu_eta_QN} are
$1/\nu_{\mathrm{CG}}=1/2$ and
$\eta_{\mathrm{CG}}=3/2$, while the exact critical moment ratio
for the SAW on the complete graph is
$Q_N^{\mathrm{CG}}=\pi/2$
~\cite{slade2020completegraph,Deng_2019}.
The LR-MF regime $0<\sigma<1/2$ is beyond the scope of the
present study. Nevertheless, an analogous coexistence of GFP
scaling and CG asymptotics has recently been studied in detail
for two-dimensional LR-percolation in its LR-MF regime~\cite{liu2026}, providing independent numerical support for
the two-scale scaling picture.

\begin{figure}[t]
    \centering
    \includegraphics[width=\linewidth]{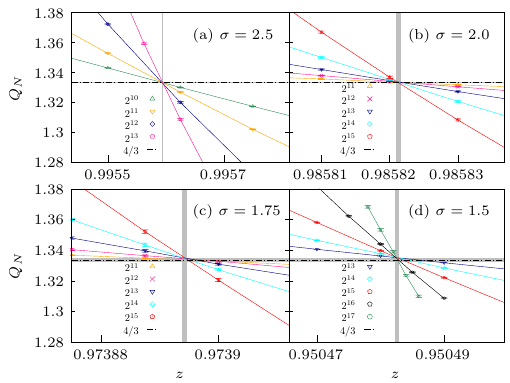}
    \caption{Binder ratio $Q_N$ for $\sigma>1$. The vertical and horizontal gray bands indicate the estimates of the critical fugacity $z_c$ and the universal critical value $Q_N^c$, respectively. For all simulated values of $\sigma$, $Q_N^c$ at criticality agrees with the SR universal value $Q_N^{\mathrm{SR}}=4/3$ within within the reported uncertainties.}
    \label{figQNabove1}
\end{figure}
\begin{figure}[htbp]
    \centering
    \includegraphics[width=0.92\linewidth]{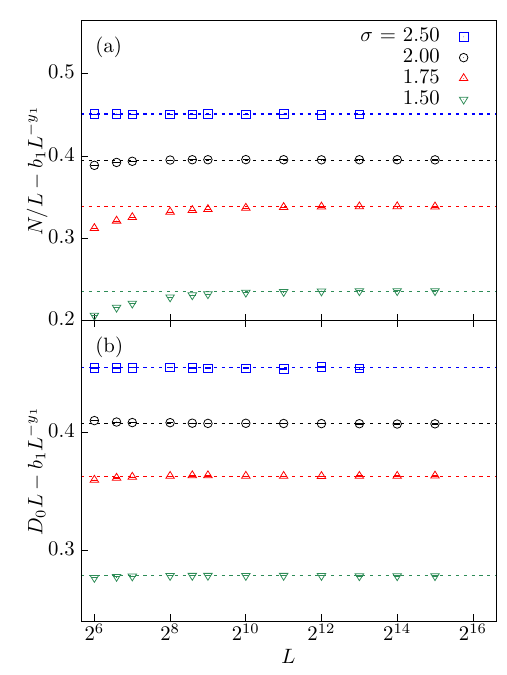}
    \caption{The semi-log plot of $N/L$ and $D_0L$ versus $L$ after subtracting the leading correction $L^{-y_1}$. The parameters $b_1$ and $y_1$ are obtained from nonlinear fitting. For $N/L$, $b_1=0.14$, $y_1=1$ at $\sigma=2.5$; $b_1=1.53$, $y_1=1$ at $\sigma=2.0$; $b_1=1.54$, $y_1=0.75$ at $\sigma=1.75$ and $b_1=1.2$, $y_1=0.53$ at $\sigma=1.5$. For $D_0L$, $b_1=0.5$, $y_1=1.0$ at $\sigma=2.5$; $b_1=0.4$, $y_1=0.76$ at $\sigma=2.0$; $b_1=0.95$, $y_1=0.75$ at $\sigma=1.75$ and $b_1=0.74$, $y_1=0.47$ at $\sigma=1.5$. Horizontal colored dashed lines are shown to guide the eye.}
    \label{fig:N_D0_above_1}
\end{figure}

\subsection{$\sigma>1$ regime}
\label{sec:results}
This subsection presents the numerical results of 1D LR-SAW. We simulate the model defined in Sec.~\ref{sec:model} using the irreversible algorithm at various values of $\sigma$. For $\sigma>1$, we simulate systems up to $L=2^{18}$ and collect more than $10^9$ samples. In this regime, the jump distribution is strongly weighted toward short distances; the resulting frequent self-avoidance rejections make the simulations particularly susceptible to critical slowing down, limiting the largest accessible system size. Statistical errors are estimated using the standard binning method and the Jackknife method.

\subsubsection{Critical fugacity $z_c$ and Binder ratio $Q_N^c$}
For $\sigma>1$, we perform large-scale simulations at $\sigma=2.5$, $2.0$, $1.75$, and $1.5$. Based on the Binder ratio $Q_N$, we examine the critical fugacity and the universal dimensionless critical quantity $Q_N^c$. Fig.~\ref{figQNabove1} shows $Q_N$ as a function of $z$ for different system sizes $L$.
The results reveal an unusual feature of the grand-canonical LR-SAW. In the regime $\sigma>1$, the critical fugacity $z_c$ differs from its NN value $z_c^{\mathrm{SR}}=1$ and the Binder ratio at criticality agrees with the SR value $Q_N^{\mathrm{SR}}=4/3$. 
The $Q_N$ curves for different $L$ exhibit well-resolved crossings at a nontrivial critical fugacity $z_c<1$, a noteworthy finite-size feature in the $T_c=0$ regime.

We estimate $z_c$ and $Q_N^c$ by performing a finite-size-scaling (FSS)
analysis of the Binder ratio $Q_N(z,L)$. For fixed $\sigma$, we define
the deviation from the critical point as
\begin{equation}
\delta z \equiv z-z_c.
\end{equation}
The thermal scaling field is an analytic function of $\delta z$ near
criticality,
\begin{equation}
\label{eq:tdef}
t(\delta z)
=\delta z+r_2 \delta z^2+O(\delta z^3),
\end{equation}
where $r_2$ is the nonuniversal coefficient of the quadratic
correction to the thermal scaling field, and the coefficient of the linear
term has been normalized to unity.
Near criticality, the Binder ratio obeys the scaling form
\begin{align}
Q_N(\delta z,L)
=\mathcal{Q}\!\left(t(\delta z,\sigma)L^{y_t},
uL^{-y_1},\ldots\right),
\label{eq:QN_scale}
\end{align}
where $u$ is the leading irrelevant scaling field, $y_1$ is the absolute
value of its RG exponent, and $y_t=1/\nu$ is the thermal scaling exponent.

To systematically estimate the critical points and universal values, we fit
the Binder ratio data to the following FSS ansatz ($y_2>y_1>0 $)
\begin{align}
\label{eq:fss_fitting_ansatz}
Q_N(\delta z,L)=&Q_N^c
+\sum_{k=1}^m a_k\left(\delta z L^{y_t}\right)^k
+b_1L^{-y_1}+b_2L^{-y_2} \nonumber\\
&+c_1\delta z L^{-y_1+y_t}
+n_2\delta z^2L^{y_t} \; .
\end{align}
Here $Q_N^c$ is the universal thermodynamic-limit value of $Q_N$ at
criticality, $m$ is the truncation order, and the $a_k$ are thermal expansion
coefficients. The coefficients $b_1$ and $b_2$ are correction amplitudes,
while $y_1$ and $y_2$ are the corresponding positive correction exponents;
throughout this work, such corrections are written as $L^{-y_k}$. The $n_2$
term originates from the nonlinear term proportional to $r_2$ in
Eq.~\eqref{eq:tdef}, while $c_1$ is the coefficient of the leading mixed
correction between the thermal and irrelevant scaling variables.

We determine $z_c$ and $Q_N^c$ through high-precision least-squares fits of the Binder ratio data to Eq.~\eqref{eq:fss_fitting_ansatz}. The fitting details and stability analyses are given in Appendix~\ref{app:fit}, and the results are summarized in Table~{\ref{table:summary}}.
The fitted results show that the critical fugacity shifts from $z_c=0.995\,593\,5(2)$ at $\sigma=2.5$ to $z_c=0.950\,482\,47(8)$ at $\sigma=1.5$. For all simulated $\sigma>1$, the fitted values of $Q_N^c$ remain consistent with the exact SR result $Q_N^{\mathrm{SR}}=4/3$ within statistical uncertainties (Fig.~\ref{figQNabove1}). 

From the phase diagram perspective, this behavior has a natural interpretation. For the 1D LR-O$(n)$ spin models with $n>0$, the $\sigma>1$ range belongs to the zero-temperature ($T_c=0$) regime. The graphical LR-SAW, however, admits a nontrivial critical fugacity whose nonuniversal value is shifted by the LR tail. This shift does not alter the universal Binder ratio: $Q_N^c$ remains consistent with $Q_N^{\mathrm{SR}}$. The $\sigma>1$ regime therefore remains in the SR universality class and belongs to the $T_c=0$ regime shown in the inset of Fig.~\ref{fig:phasediag}.

\subsubsection{Critical exponents $\nu$ and $\eta$ }
We next investigate the correlation-length exponent $\nu$ and the anomalous dimension $\eta$ in the regime $\sigma>1$. At the critical fugacity determined above, we perform large-scale simulations for each value of $\sigma$. The exponents are extracted from the FSS of the mean walk length $N$ and the empty-walk probability $D_0$, respectively. At criticality, the mean walk length obeys the scaling form
\begin{equation}
\label{eqn}
N = L^{1/\nu}\left(a+b_1L^{-y_1}+b_2L^{-y_2}\right)+c,
\end{equation}
where $y_2>y_1>0$ are correction exponents, $a$, $b_1$, and $b_2$ are nonuniversal amplitudes, and $c$ accounts for the analytic background.
We perform least-squares fits of the data to the FSS form in Eq.~\eqref{eqn}. The resulting exponent estimates are summarized in Table~\ref{table:summary}, and the fitting details are given in Appendix~\ref{app:fit}.

For $\sigma>1$, the semi-log plot of $N/L$ versus $L$ is shown in Fig.~\ref{fig:N_D0_above_1}(a) after subtracting the leading correction $L^{-y_1}$. All data for $\sigma>1$ exhibit a clear horizontal trend in semi-log coordinates at large $L$. This means that the numerical estimates of $1/\nu$ are fully consistent with the SR value $1/\nu_{\rm SR}=1$.

We then estimate $\eta$ from the empty-walk probability $D_0$. At criticality, the FSS takes the form
\begin{equation}
D_0 = L^{\eta-2}(a+b_1L^{-y_1}+b_2L^{-y_2}).
\label{d0eq}
\end{equation}
Here $y_2>y_1>0$ are correction exponents, and $a$, $b_1$, and $b_2$ are nonuniversal amplitudes. For $\sigma > 1$, Fig.~\ref{fig:N_D0_above_1}(b) shows that $D_0L$ approaches an asymptotic constant for all $\sigma>1$, implying $D_0\sim L^{-1}$ and $\eta=1$. The numerical estimates are consistent with $\eta=\eta_{\rm SR}=1$ within statistical errors.

Taken together with the Binder ratio results, the thermal estimate $1/\nu=1$ and the magnetic estimate $\eta=1$ provide mutually consistent evidence that the asymptotic critical behavior for $\sigma>1$ is governed by the SR fixed point. The LR tail shifts the nonuniversal critical fugacity and produces clear finite-size crossings of $Q_N$, but it does not change the universal dimensionless quantity or the thermal and magnetic critical exponents within our numerical resolution. This critical behavior is consistent with that of the SR case, belongs to the SR universality class, and lies in the $T_c=0$ regime of the LR-O$(n)$ universality diagram.

\begin{figure}[t]
    \centering
    \includegraphics[width=1.0\linewidth]{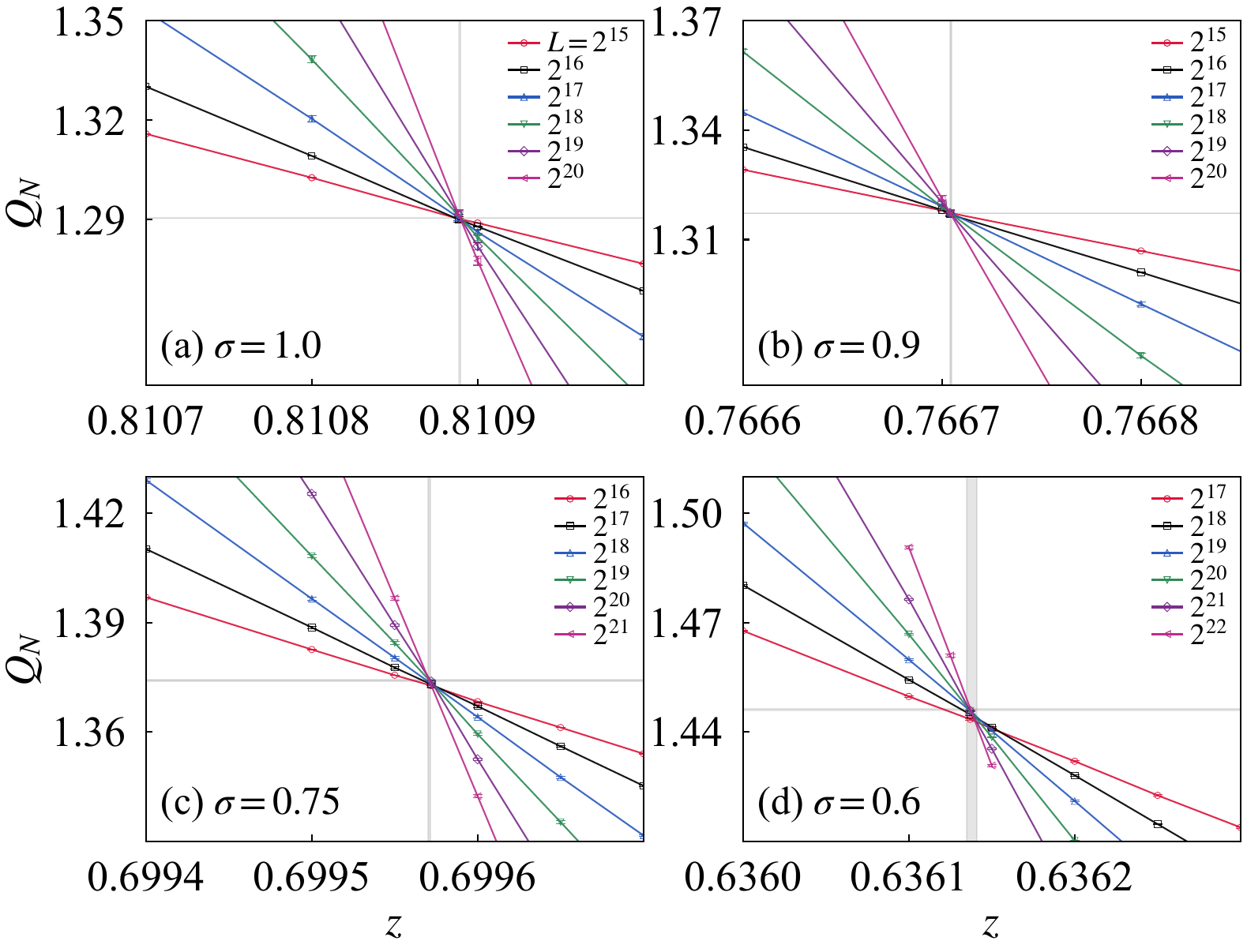}
    \caption{Binder ratio $Q_N$ for $\sigma \le 1$. The vertical and horizontal gray bands indicate the estimated values of $z_c$ and $Q_N^c$ from FSS, respectively. 
    The Binder ratio $Q_N(z,L)$ with different size exhibit a clear crossing 
    in the vicinity of critical fugacity $z_c$.}
    \label{figqnplot}
\end{figure}

\subsection{$1/2<\sigma \le 1$ regime}

For $1/2<\sigma\le1$, we simulate $\sigma=1.0$, $0.9$, $0.75$, and
$0.6$. The irreversible algorithm~\cite{hu2017irreversible} allows us to reach maximum system sizes
of $L=2^{20}$, $2^{20}$, $2^{21}$, and $2^{22}$, respectively, with more
than $10^7$ samples collected in each case.

\subsubsection{Critical fugacity $z_c$ and Binder ratio $Q_N^c$}
Using the same fitting framework as in the $\sigma>1$ regime, we determine
$z_c$ and $Q_N^c$ through high-precision least-squares multiparameter fits
of the Binder ratio data to Eq.~\eqref{eq:fss_fitting_ansatz}. The fitting details and stability analyses are given in Appendix~\ref{app:fit}, and the fitting results are summarized in Table~\ref{table:summary}.

The critical fugacity decreases monotonically as $\sigma$ decreases, as shown in the main panel of Fig.~\ref{fig:phasediag}.
At $\sigma=1$, the fit gives $Q_N^c=1.290\,57(2)$, in clear contrast to the SR value $Q_N^{\mathrm{SR}}=4/3$ found throughout the simulated $\sigma>1$ regime (Fig.~\ref{fig:nu_eta_QN}(c)). This sharp change provides strong evidence that the universality class changes at $\sigma=1$: the SR universality observed for $\sigma>1$ gives way to LR universality for $\sigma\le1$, away from the $T_c=0$ regime. Similar discontinuous changes of universal quantities at LR--SR crossovers have also been observed in other LR systems~\cite{liu2025twodimensionalpercolationmodellongrange,xiao2025sakscriterionstatisticalmodels}. For $\sigma<1$, the fitted value of $Q_N^c$ increases monotonically as $\sigma$ decreases. The point $\sigma=1/2$ separates the LR-WF-B and LR-MF regimes, where $Q_N^c$ is expected to reach the complete-graph (CG) with value $Q_N^{\mathrm{CG}}=\pi/2$~\cite{Deng_2019,slade2020completegraph}. The nature of this boundary and the LR-MF regime are beyond the scope of the present work and are not investigated here.

As an additional consistency check of the full FSS fits, we extrapolate the crossings of the Binder ratio curves. For each pair of adjacent system sizes $(L,2L)$, we impose the crossing condition
\begin{equation}
    Q_N(z_c(L),L) = Q_N(z_c(L),2L),
\end{equation}
which determines a size-dependent crossing point $z_c(L)$ and the corresponding Binder ratio 
$Q_N(z_c(L), L)$. Operationally, these crossing coordinates are obtained by interpolating the discrete numerical data for each $L$ with cubic splines and locating the intersection of the $L$ and $2L$ curves via bisection.
The leading FSS forms for these crossing sequences are~\cite{liu2025twodimensionalpercolationmodellongrange}

\begin{align}
    z_c(L) &= z_c + a L^{-(y_t+y_1)} + \cdots, \nonumber\\
    Q_N(z_c(L),L) &= Q_N^c + b L^{-y_1} + \cdots.
\end{align}
Here $a$ and $b$ are nonuniversal crossing amplitudes, $y_1>0$ is the
leading correction-to-scaling exponent defined above, and the ellipses
denote subleading corrections. Figure \ref{figqnplot} shows the Binder ratio $Q_N(z,L)$ for different system sizes, together with the critical fugacity $z_c$ and the critical Binder ratio $Q^c_N$ obtained from the multiparamete fits. The Binder-ratio curves exhibit a pronounced crossing behavior in the vicinity of $z_c$,
providing a consistency check with the FSS analysis. 

\begin{figure}[t]
     \centering
     \includegraphics[width=1.0\linewidth]{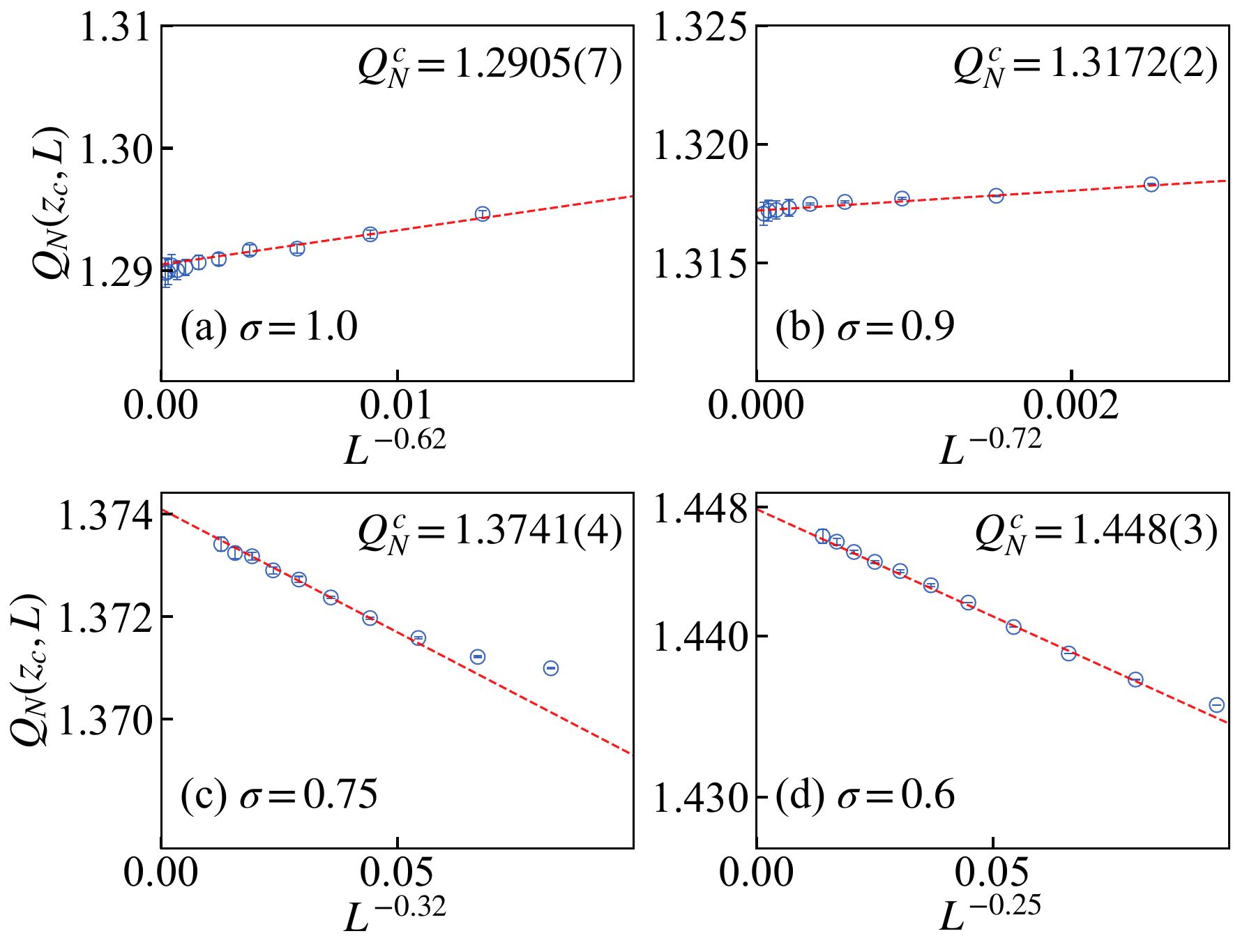}
     \caption{The asymptotic linear dependence of $Q_N(z_c,L)$ on $L^{-y_1}$ provides a consistency check of the leading correction form. The intercept of the fitted line yields an additional consistency check of $Q_N^c$.}
     \label{fig:qny1}
 \end{figure}

Using the extrapolated estimate of $z_c$, we evaluate $Q_N(z_c,L)$
for each $L$ from a local quadratic fit,
\begin{equation}
    Q_N(z, L) = Q_N(z_c, L) + a_1 (z - z_c) + a_2 (z- z_c)^2,
\end{equation}
where $a_1$ and $a_2$ are fitted separately for each $L$.
We then fit the size dependence to the critical-point form of
Eq.~\eqref{eq:fss_fitting_ansatz},
\begin{equation}
    Q_N(z_c, L) = Q_N^c + b_1 L^{-y_1} + b_2 L^{-y_2}.
\end{equation}
For sufficiently large $L$, the subleading term becomes negligible,
so $Q_N(z_c,L)$ approaches a straight line when plotted against
$L^{-y_1}$, with intercept $Q_N^c$. 
With the appropriate value of $y_1$, the asymptotic behavior of the extrapolated $Q_N(z_c,L)$ versus $L^{-y_1}$ is shown in Fig.~\ref{fig:qny1}. 
The line intercept yields an additional consistency check of $Q_N^c$, 
which is consistent with the values reported in Table~\ref{table:summary}, providing an additional consistency check. Here, the uncertainty associated with the crossing-point extrapolation of $z_c$ is larger than that obtained from the multiparameter FSS fit, and we use the crossing-based estimate of $z_c$ only as an intermediate input for extrapolating $Q_N(z_c,L)$, rather than reporting it as an independent estimate. 

Taken together, these results show that throughout $1/2<\sigma\le1$ the
critical fugacity varies smoothly with $\sigma$, whereas the critical Binder
ratio exhibits LR universal behavior distinct from the SR universality
class. In particular, $Q_N^c(\sigma=1)=1.290\,57(2)$ differs clearly from
$Q_N^{\mathrm{SR}}=4/3$. As $\sigma$ decreases below $1$, $Q_N^c$
increases monotonically toward the expected CG limiting value
$Q_N^{\mathrm{CG}}=\pi/2$ at $\sigma=1/2$
~\cite{Deng_2019,slade2020completegraph}. At $\sigma=1$, the fitted
value of $Q_N^c$, together with the finite-size behavior shown in
Fig.~\ref{figqnplot}, indicates a conventional continuous transition
rather than the BKT-like transition of the 1D LR-Ising model at the
same boundary~\cite{frohlich1982,PhysRevLett.37.1577}. The estimates
obtained by interpolation and extrapolation agree closely with those from
the full multiparameter fits and serve only as an additional consistency check. The final estimates of $Q_N^c$ reported in
Table~\ref{table:summary} are obtained from the fits to
Eq.~\eqref{eq:fss_fitting_ansatz}, with the fitting details given in Appendix~\ref{app:fit}.

\begin{figure}[t]
    \centering
    \includegraphics[width=0.9\linewidth]{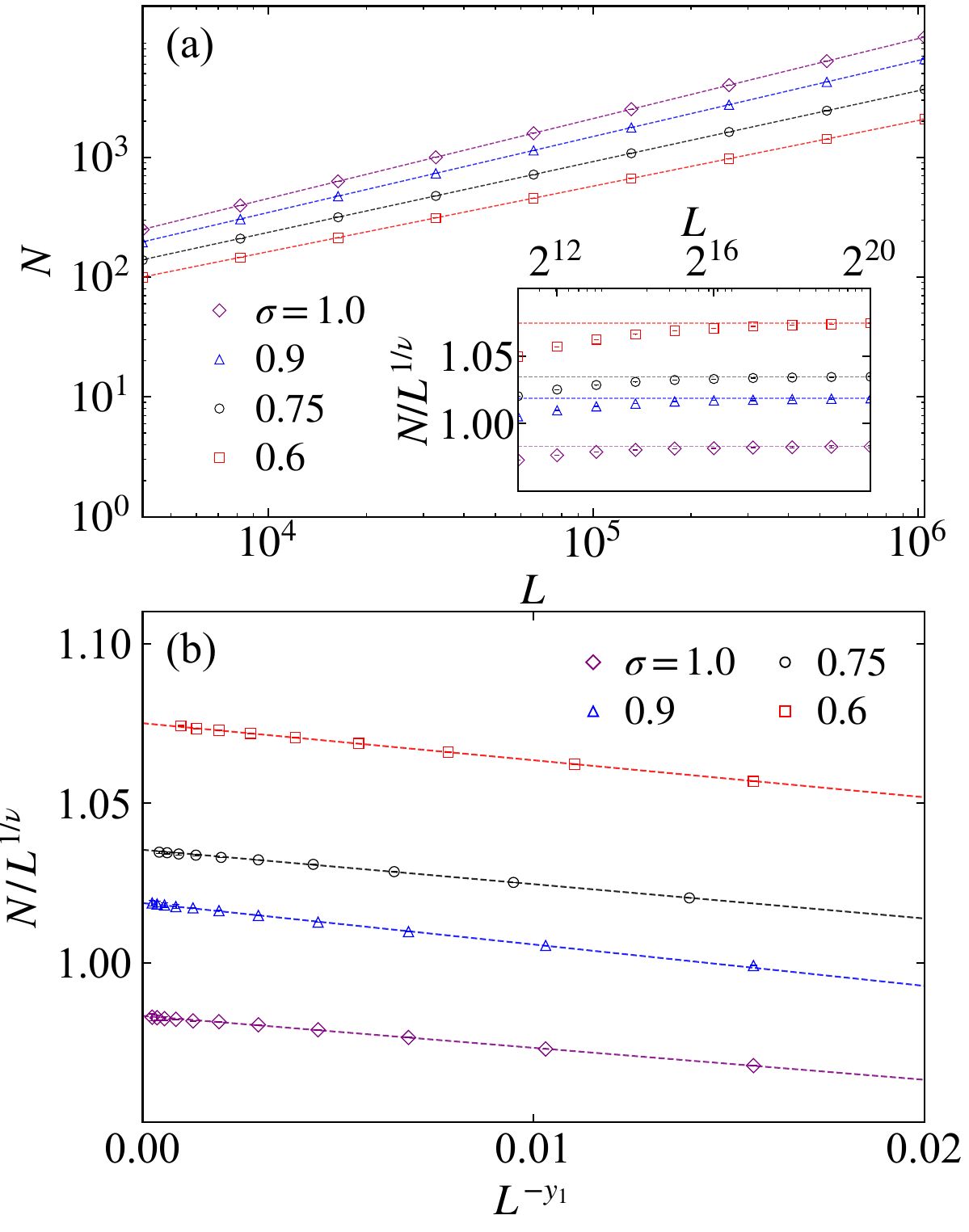}
    \caption{(a) Log-log plot of $N$ at criticality versus the system size
    $L$ for different values of $\sigma$. The dashed lines illustrate the
    fitted leading behavior $N\sim L^{1/\nu}$. The inset shows that
    $N/L^{1/\nu}$ approaches a finite constant as $L$ increases. (b)
    Asymptotically linear dependence of $N/L^{1/\nu}$ on $L^{-y_1}$, using
    $y_1=0.6$ for $\sigma=1.0$ and $0.9$; $y_1=0.56$ for $\sigma=0.75$;
    and $y_1=0.5$ for $\sigma=0.6$.}
    \label{fitn}
\end{figure}
\subsubsection{Critical exponents $\nu$ and $\eta$}
All exponent fits in this subsection are performed at the critical
fugacities $z_c(\sigma)$ determined above. For $1/2<\sigma\le1$, we extract
$1/\nu$ from the finite-size scaling of $N$ and $\eta$ from that of $D_0$,
using Eqs.~\eqref{eqn} and~\eqref{d0eq}, respectively, following the same
procedure as for $\sigma>1$. The estimates are summarized in
Table~\ref{table:summary}, the fitting details are given in
Appendix~\ref{app:fit}, and the results are displayed in
Figs.~\ref{fig:nu_eta_QN}(a) and (b).

At criticality, the mean walk length obeys $N\sim L^{1/\nu}$, as shown
in Fig.~\ref{fitn}(a). Using the fitted values of $1/\nu$ and $y_1$, the
rescaled quantity $N/L^{1/\nu}$ approaches a finite constant as $L$
increases [inset of Fig.~\ref{fitn}(a)] and exhibits an asymptotically
linear dependence on $L^{-y_1}$ [Fig.~\ref{fitn}(b)], consistent with the
reduced leading-correction form $N/L^{1/\nu}=a+b_1L^{-y_1}$ of
Eq.~\eqref{eqn}. At $\sigma=1$, we obtain $1/\nu=0.666\,2(4)$, clearly
different from the SR value $1/\nu_{\mathrm{SR}}=1$. This finite value
implies conventional power-law critical scaling at the boundary, in
contrast to the BKT-like criticality of the 1D LR-Ising model, for which
$1/\nu\to0$~\cite{frohlich1982,PhysRevLett.37.1577,Pagni2025}. As
$\sigma$ decreases below $1$, $1/\nu$ decreases monotonically, as shown in
Fig.~\ref{fig:nu_eta_QN}(a). The discontinuity in $1/\nu$ at $\sigma=1$ and its subsequent
variation provide clear evidence for the change from SR to LR
universality.

An instructive comparison is provided by the Flory-type approximation for SAW. For the SR-SAW, balancing the entropic and excluded-volume contributions gives $\nu_{\mathrm{F}}^{\mathrm{SR}}=3/(d+2)$, or $1/\nu_{\mathrm{F}}^{\mathrm{SR}}=(d+2)/3$. Its LR extension predicts
\begin{equation}
    \frac{1}{\nu_{\rm F}}=\frac{d+\sigma}{3}.
\end{equation}
For $d=1$, this gives $1/\nu_{\mathrm{F}}=2/3$ at $\sigma=1$
~\cite{de1979scaling,halley1985node,grassberger1985critical}. Since the LR
tail is irrelevant for $\sigma>1$, we restrict this comparison to
$1/2<\sigma\le1$. Fig.~\ref{fig:nu_eta_QN}(a) displays the deviation
$\delta_{1/\nu}=1/\nu-1/\nu_{\mathrm{F}}$. At $\sigma=0.6$, we find
$\delta_{1/\nu}\simeq0.0126$, and the deviations at $\sigma=0.6$ and
$0.75$ are statistically resolvable. By contrast, at $\sigma=0.9$ and
$1.0$, the estimates are consistent with the LR Flory prediction within
the numerical uncertainties. Thus, the Flory expression provides a good
numerical approximation to the thermal exponent near $\sigma=1$, although
the resolved deviations show that it is not exact throughout the LR regime.

\begin{figure}[t]
    \centering
    \includegraphics[width=0.9\linewidth]{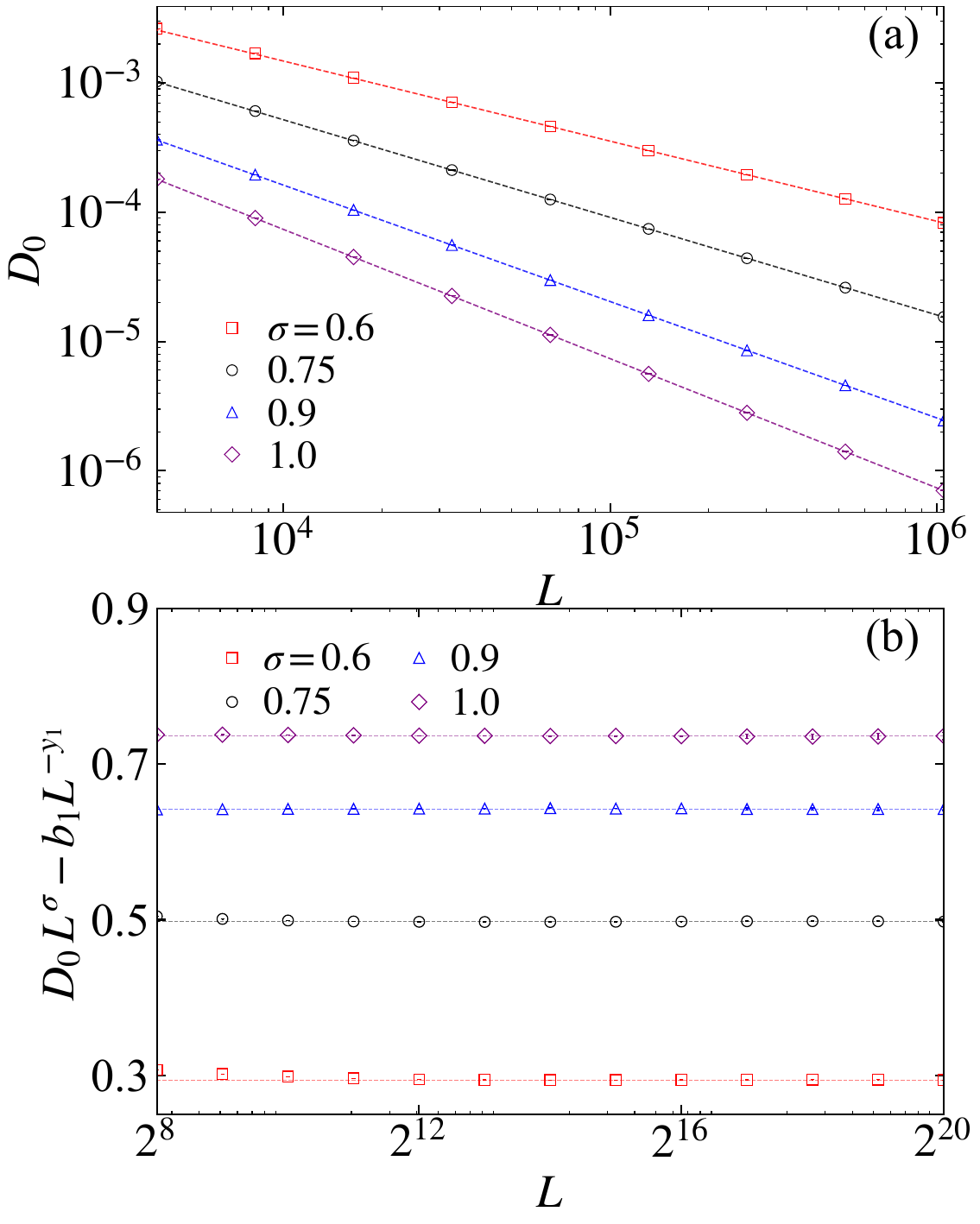}
    \caption{(a) Log-log plot of $D_0$ versus system size $L$ for different values of $\sigma$. The dashed lines show power-law fits to
$D_0\sim L^{\eta-2}$, with $\eta$ treated as a fitting parameter. (b) The semi-log plot of $D_0L^{\sigma}$ versus $L$ after subtracting the leading correction $L^{-y_1}$, using $y_1 = 0.45$ for $\sigma = 1.0$; $y_1 = 0.41$ for $\sigma = 0.9$; $y_1 = 0.2$ for $\sigma = 0.75$; and $y_1 = 0.14$ for $\sigma = 0.6$. }
    \label{figdplot}
\end{figure}
We determine the anomalous dimension $\eta$ by fitting $D_0$ to
Eq.~\eqref{d0eq}.
Fig.~\ref{figdplot}(a) shows the log-log plots of $D_0$ versus $L$
together with the corresponding fits. The resulting estimates agree with
the LR-GFP value $\eta_{\mathrm{GFP}}=2-\sigma$ for all simulated
$\sigma\le1$. To display this agreement and the corrections to scaling,
Fig.~\ref{figdplot}(b) shows $D_0L^\sigma$ after subtracting the
fitted leading correction $b_1L^{-y_1}$. The corrected data approach
constants at large $L$, as expected from the reduced form
$D_0L^\sigma=a+b_1L^{-y_1}$ of Eq.~\eqref{d0eq}. At $\sigma=1$, we obtain
$\eta=0.999\,7(4)$, which agrees with the SR value
$\eta_{\mathrm{SR}}=1$ within one standard deviation. Thus, unlike
$1/\nu$, $\eta$ connects continuously to its SR value at the boundary.
The inset of Fig.~\ref{fig:nu_eta_QN}(b) shows
$\delta_\eta=\eta-\eta_{\mathrm{GFP}}$, which is consistent with zero
within one standard deviation for every simulated $\sigma\le1$, supporting
$\eta=\eta_{\mathrm{GFP}}=2-\sigma$ throughout the simulated LR regime,
including $\sigma=1$.

Taken together, throughout $1/2<\sigma\le1$, $Q_N^c$ and $1/\nu$ differ
from their SR values, whereas the anomalous dimension follows
$\eta=2-\sigma$. At $\sigma=1$, the discontinuous changes in $Q_N^c$ and
$1/\nu$ contrast with the continuous matching of $\eta$ to
$\eta_{\mathrm{SR}}=1$. Moreover, $\nu\ne\nu_{\mathrm{GFP}}$ but
$\eta=\eta_{\mathrm{GFP}}$, which is the characteristic exponent structure
of the LR-WF-B regime
~\cite{xiao2025universalitydiagramphasetransitions}. These combined results
identify $\sigma=1$ as the boundary at which the model changes from the SR
universality class in the $T_c=0$ regime to the LR-WF-B regime, consistent
with the inset of Fig.~\ref{fig:phasediag}.

\section{Conclusion}
\label{sec:conclusion}
We have studied the 1D LR-SAW in a finite-volume grand-canonical ensemble using large-scale Monte Carlo simulations. From the finite-size scaling of the Binder ratio, the mean walk length, and the empty-walk probability, we obtained high-precision estimates of the critical fugacity $z_c$, the universal Binder ratio $Q_N^c$, and the exponents $1/\nu$ and $\eta$. The agreement between the Binder-ratio crossing extrapolations and
the full FSS fits, together with the stability of the exponent fits,
supports the robustness of the numerical results.

Our results identify $\sigma = 1$ as the boundary between SR and LR universality. For $\sigma > 1$, the long-range jump distribution lowers the nonuniversal critical fugacity relative to the NN limit, while the universal quantities remain consistent with the exact SR-SAW values: $Q_N^c = 4/3$, $1/\nu = 1$, and $\eta = 1$. Thus, the asymptotic critical behavior in this regime is governed by the SR endpoint, or $T_c = 0$, universality class despite the algebraic tail in the jump distribution.

At $\sigma = 1$, in contrast to the LR-Ising model with a BKT-like transition, the LR-SAW undergoes a conventional continuous transition with
power-law critical scaling. The measured values $Q_N^c = 1.290\,57(2)$ and $1/\nu = 0.666\,2(4)$ differ clearly from their SR counterparts, whereas $\eta \simeq 1$ is consistent with both the exact SR value and the LR relation $\eta = 2 - \sigma$. Hence, $\eta$ varies smoothly across the boundary, while $Q_N^c$ and $1/\nu$ exhibit a sharp change.
The value $1/\nu=0.666\,2(4)$ is also consistent with the
Flory-type estimate $(d+\sigma)/3=2/3$.

For $1/2 < \sigma < 1$, $Q_N^c$ and $1/\nu$ vary nontrivially with $\sigma$. The anomalous dimension follows the LR Gaussian fixed-point prediction $\eta = 2 - \sigma$ with high numerical accuracy, whereas $1/\nu$ is not described by the LR Gaussian relation and agrees numerically with the Flory-type estimate near $\sigma=1$
but exhibits statistically resolved deviations at $\sigma=0.75$
and $0.6$. This combination of $\eta=\eta_{\mathrm{GFP}}$ and
$\nu\ne\nu_{\mathrm{GFP}}$ is consistent with the proposed LR-WF-B regime of LR-O$(n)$ models~\cite{xiao2025universalitydiagramphasetransitions}. Overall, our high-precision results provide a useful benchmark for future analytical and field-theoretic studies of LR systems.

Recent CFT work on the 1D LR-Ising model has proposed a dual
description that becomes weakly coupled near $\sigma=1$, enabling
perturbative calculations of CFT data~\cite{Benedetti2025}.
This advance motivates the search for analogous analytical and
conformal descriptions across the broader family of LR-O$(n)$ models.
As a high-precision realization of the $n\to0$ limit, the LR-SAW
results reported here provide useful benchmarks for future
RG, conformal-bootstrap, and duality-based studies,
and may contribute to a unified understanding of LR criticality and
the LR--SR crossover across different values of $n$.


\section*{Acknowledgments}
We acknowledge the support from the National Natural Science Foundation of China (NSFC) under Grant No. 12275263, as well as Quantum Science
and Technology-National Science and Technology Major
Project (under Grant No. 2021ZD0301900). ZF is also
supported by the National Natural Science Foundation
of China (NSFC) under Grant No. 12504265.

\bibliography{ref}
\clearpage

\appendix
\section{Exact results for 1D SR-SAW}
\label{app:sr_exact}
Here we derive the exact critical fugacity and critical exponents of the 1D NN SAW in the finite-volume grand-canonical ensemble used in this work. Consider a periodic chain of length $L$ with the walk rooted at the origin. For a nonempty walk, once the direction of the first step is chosen, self-avoidance forces all subsequent steps to continue in the same direction. The number of allowed configurations is therefore
\begin{equation}
C_N=
\begin{cases}
1, & N=0,\\
2, & 1\le N\le L-1.
\end{cases}
\end{equation}
The finite-volume grand-canonical partition function is then
\begin{equation}
\mathcal{Z}_L^{\mathrm{SR}}(z)=1+2\sum_{N=1}^{L-1} z^N.
\end{equation}
For $z\neq 1$, this can be written as
\begin{equation}
\mathcal{Z}_L^{\mathrm{SR}}(z)=1+\frac{2z\left(1-z^{L-1}\right)}{1-z}.
\end{equation}
In the thermodynamic limit, the series converges for $z<1$ and becomes singular at $z=1$. Hence the exact critical fugacity is
\begin{equation}
z_c^{\mathrm{SR}}=1.
\end{equation}

At criticality, the mean walk length is
\begin{equation}
N_{\mathrm{SR}}
=\frac{2\sum_{N=1}^{L-1}N}{2L-1}
=\frac{L(L-1)}{2L-1}
=\frac{L}{2}+O(1),
\end{equation}
which implies
\begin{equation}
1/\nu_{\mathrm{SR}}=1.
\end{equation}
The empty-walk probability at criticality is
\begin{equation}
D_{0}^{\mathrm{SR}}
=\frac{1}{\mathcal{Z}_L^{\mathrm{SR}}(1)}
=\frac{1}{2L-1}
=\frac{1}{2}L^{-1}\left[1+O(L^{-1})\right].
\end{equation}
Using the FSS relation $D_0\sim L^{\eta-2}$, we obtain $\eta_{\rm SR}=1$.
Equivalently, let $G_L(r;z)$ denote the unnormalized end-to-end
two-point function introduced in the main text.  At $z=1$, each
nonzero endpoint can be reached by two monotone walks around the ring,
and hence
\begin{equation}
G_L(r;1)=2.
\end{equation}
After normalization over all endpoint positions, the corresponding
endpoint probability is
\begin{equation}
\frac{G_L(r;1)}{\mathcal Z_L^{\rm SR}(1)}
=\frac{2}{2L-1}\asymp L^{-1},
\end{equation}
which gives the $L$-dependent plateau described in the Introduction.
For fixed separation $r$ in the thermodynamic limit, the winding
contribution vanishes for $z<1$, and the two-point function becomes
\begin{equation}
G(r;z)=z^{|r|}=e^{-|r|/\xi},
\end{equation}
where $\xi=-1/\ln z$.  As $z\to z_c^-=1$,
$\xi\sim(1-z)^{-1}=(z_c-z)^{-1}$, so comparison with
$\xi\sim(z_c-z)^{-\nu}$ gives $\nu_{\rm SR}=1$.
At criticality, $G(r;z_c)\asymp1$.  Comparing this result with
$G(r;z_c)\sim r^{2-d-\eta}=r^{1-\eta}$ in $d=1$ again gives
$\eta_{\rm SR}=1$.
Hence, the exact values for the 1D NN-SAW are
\begin{equation}
z_c^{\mathrm{SR}}=1, \quad \nu_{\rm SR}=1, \quad \eta_{\rm SR}=1.
\end{equation}

\section{Fitting strategy and detailed fits of the observables}
\label{app:fit}
In this appendix, we describe the fitting procedure used to extract the critical fugacity $z_c$, the universal Binder ratio $Q_N^c$, and the critical exponents $1/\nu$ and $\eta$. All fits are based on the finite-size scaling ansatz in Eq.~\eqref{eq:fss_fitting_ansatz} and its analogues for $N$ and $D_0$, which include a leading correction exponent $y_1>0$ and a subleading one $y_2>y_1$, and, where appropriate, a coupling term between 
the thermal and irrelevant scaling fields.

For each $\sigma$, we repeat the fits with increasing
$L_{\min}$ and monitor the reduced chi-square $\chi^2/\mathrm{DF}$ and the stability
of $z_c$, $Q_N^c$, $\nu$, and $\eta$.
Here $\chi^2$ denotes the weighted sum of squared residuals, and DF (degrees of freedom) is the number of fitted data points minus the number of free fit parameters.
With reliable error estimates, $\chi^2/\mathrm{DF}$ is expected to be close to unity for a statistically consistent fit.
We retain fitting windows with acceptable goodness of fit
and estimates consistent within uncertainties as
$L_{\min}$ is further increased.

An important step in the fitting analysis is to test whether the leading correction-to-scaling exponent is independently resolved by the data. In the full multi-parameter fits used to extract $z_c$, $Q_N^c$ and the critical exponents, we perform both free-$y_1$ and fixed-$y_1$ analyzes to test the stability of the resulting estimates. We also check whether the free-fit estimates of $y_1$ stabilize within the asymptotic fitting window and agree with the value inferred from the large-$L$ linear approach of $Q_N(z_c,L)$ versus $L^{-y_1}$ in Fig.~\ref{fig:qny1}. Together, these tests confirm the mutual consistency of the full FSS analysis, the large-$L$ correction analysis, and the cross-extrapolation procedure.

We also test the nonlinear thermal-field term proportional to $n_2$ in
Eq.~\eqref{eq:fss_fitting_ansatz}. Among the values of $\sigma$ studied,
a statistically resolvable contribution $n_2$ is observed only at
$\sigma=2.5$. For all other values of $\sigma$, the coefficient cannot be
resolved within a stable fitting window, and therefore the term $n_2$ is
omitted from the corresponding fits reported in the following.

For the region $\sigma>1$, detailed fitting results for $Q_N$, $N$, and $D_0$ are presented in Tables \ref{qn_large}, \ref{n_large}, and \ref{d_large}, respectively. For the region $1/2<\sigma\leq 1$, the corresponding fitting results are given in Tables \ref{tabqn}, \ref{tabn}, and \ref{tabd}, respectively.

\begin{table*}
\footnotesize
\caption{Results of nonlinear fits of the Binder ratio $Q_N(z,L)$ to Eq.~\eqref{eq:fss_fitting_ansatz} for
$\sigma=2.5$, $2.0$, $1.75$, $1.5$. Values listed without uncertainties are held fixed, whereas ``-'' indicates that the corresponding term is omitted. }
\label{qn_large}
\begin{tabular*}{\textwidth}{@{\extracolsep{\fill}}llllllllllllll}
\hline\hline
$\sigma$ & $L_{\min}$ & $\chi^2$/DF & $z_c$ & $Q_N^c$ & $y_t(1/\nu)$ & $a_1$ & $a_2$ & $b_1$ & $b_2$ & $y_1$ & $y_2$ & $c_1$ & $n_2$ \\ \hline
2.5 & 64 & 45.1/47 & 0.9955936(2) & 1.33324(8) & 1.0038(8) & -0.0965(5) & 0.0080(3) & -0.5(1) & - & 1.18(5) & - & - & - \\
 & 48 & 91.0/53 & 0.9955932(3) & 1.3334(1) & 1 & -0.0990(2) & 0.0087(3) & -0.25(4) & - & 1.02(4) & - & - & - \\
 & 64 & 67.8/48 & 0.9955936(2) & 1.33322(9) & 1 & -0.0990(1) & 0.0087(3) & -0.5(1) & - & 1.20(6) & - & - & - \\
 & 512 & 10.7/21 & 0.9955941(1) & 1.33294(4) & 1 & -0.0995(1) & 0.0085(2) & - & - & - & - & - & - \\
 & 1024 & 9.8/16 & 0.9955941(2) & 1.33299(8) & 1 & -0.0995(1) & 0.0084(2) & - & - & - & - & - & - \\
 & 48 & 39.8/51 & 0.9955934(2) & 1.33344(7) & 1 & -0.0997(2) & 0.0083(3) & -0.23(2) & - & 1.00(3) & - & 0.19(4) & 0.5(1) \\
 & 64 & 25.7/46 & 0.9955937(2) & 1.33327(6) & 1 & -0.0996(1) & 0.0083(2) & -0.41(6) & - & 1.14(4) & - & 0.38(9) & 0.4(1) \\
 & 8 & 96.6/75 & 0.9955933(2) & 1.33348(4) & 1 & -0.0995(2) & 0.0082(3) & -1.6(1) & 2.5(1) & 1.26(1) & 1.5 & 0.23(6) & 0.6(2) \\
 & 16 & 42.5/65 & 0.9955935(2) & 1.33339(5) & 1 & -0.0996(1) & 0.0082(2) & -2.1(4) & 3.0(5) & 1.30(3) & 1.5 & 0.51(9) & 0.6(1) \\
 & 24 & 39.8/60 & 0.9955934(2) & 1.33341(6) & 1 & -0.0996(1) & 0.0082(2) & -1.7(7) & 2.5(9) & 1.27(6) & 1.5 & 0.6(2) & 0.5(1) \\
 & 32 & 34.0/55 & 0.9955934(2) & 1.33341(7) & 1 & -0.0997(1) & 0.0083(2) & -3(2) & 4(3) & 1.3(1) & 1.5 & 0.8(3) & 0.5(1) \\
\hline
2.0 & 96  & 11.8/50 & 0.98582105(2) & 1.33363(5) & 0.999(4) & -0.090(4) & 0.007(1) & -0.86(7) & - & 1.07(2) & - & - & - \\
 & 128 & 9.7/45  & 0.98582111(2) & 1.33353(5) & 0.998(4) & -0.091(4) & 0.007(1) & -1.2(2)  & - & 1.14(3) & - & - & - \\ 
 & 64  & 36.7/57 & 0.98582134(4) & 4/3 & 1 & -0.0890(4) & 0.006(1) & -0.87(3) & - & 1.083(9) & - & - & - \\
 & 96  & 18.2/52 & 0.98582126(3) & 4/3 & 1 & -0.0890(3) & 0.007(1) & -1.31(9) & - & 1.17(1) & - & - & - \\ 
 & 96  & 78.9/53 & 0.98582133(5) & 4/3 & 1 & -0.0889(7) & 0.007(2) & -0.595(4) & - & 1 & - & - & - \\
 & 128 & 60.9/48 & 0.98582132(5) & 4/3 & 1 & -0.0889(6) & 0.007(2) & -0.576(6) & - & 1 & - & - & - \\  
 & 64  & 14.6/55 & 0.98582098(2) & 1.33376(4) & 1 & -0.0892(3) & 0.0074(9) & -0.64(2) & - & 1.00(1) & - & 1.0(7) & - \\
 & 64  & 14.6/56 & 0.98582098(2) & 1.33376(2) & 1 & -0.0892(3) & 0.0074(8) & -0.643(2) & - & 1 & - & 1.0(7) & - \\ 
 & 64  & 15.2/56 & 0.98582098(2) & 1.33376(4) & 1 & -0.0891(3) & 0.0074(9) & -0.64(2) & - & 1.00(1) & - & - & - \\
 & 96  & 11.8/51 & 0.98582105(2) & 1.33363(5) & 1 & -0.0891(3) & 0.0073(8) & -0.86(7) & - & 1.07(2) & - & - & - \\  
 & 512  & 51.5/33 & 0.98582156(6) & 1.33287(8) & 1 & -0.0891(7) & 0.006(2) & - & - & - & - & - & - \\
 & 1024 & 18.0/28 & 0.98582127(4) & 1.33325(7) & 1 & -0.0891(4) & 0.007(1) & - & - & - & - & - & - \\  
 & 2048 & 22.9/23 & 0.9858216(1) & 1.3333(1) & 1 & -0.0893(5) & - & - & - & - & - & - & - \\
 & 4096 & 22.1/17 & 0.9858216(1) & 1.3334(2) & 1 & -0.0893(6) & - & - & - & - & - & - & - \\
\hline
1.75 
 & 96 & 59.3/49 & 0.9738944(2) & 1.3355(2) & 1 & -0.0761(8) & - & -0.34(4) & -0.4(1) & 0.75 & 1 & 0.1(4) & - \\
 & 128 & 31.3/44 & 0.9738948(2) & 1.3348(2) & 1 & -0.0762(6) & - & -0.13(5) & -1.1(1) & 0.75 & 1 & 0.2(3) & - \\
 & 256 & 20.9/39 & 0.9738951(2) & 1.3341(2) & 1 & -0.0762(5) & - & 0.24(9) & -2.4(3) & 0.75 & 1 & 0.1(4) & - \\ 
 & 128 & 31.5/45 & 0.9738948(2) & 1.3348(2) & 1 & -0.0760(5) & - & -0.13(5) & -1.1(1) & 0.75 & 1 & - & - \\
 & 256 & 21.0/40 & 0.9738951(2) & 1.3341(2) & 1 & -0.0761(5) & - & 0.24(9) & -2.4(3) & 0.75 & 1 & - & - \\
 & 512 & 19.6/30 & 0.9738951(3) & 1.3342(5) & 1 & -0.0761(5) & - & 0.1(3) & -2(1) & 0.75 & 1 & - & - \\ 
 & 2048 & 17.1/21 & 0.9738948(3) & 1.3350(5) & 1 & -0.0761(6) & - & -0.4(2) & - & 0.75 & - & - & - \\
 & 4096 & 16.1/16 & 0.9738950(5) & 1.334(1) & 1 & -0.0761(6) & - & -0.1(6) & - & 0.75 & - & - & - \\
 & 8192 & 15.0/11 & 0.973896(1) & 1.332(3) & 1 & -0.0761(8) & - & 2(2) & - & 0.75 & - & - & - \\ 
 & 1024 & 26.3/26 & 0.9738952(1) & 1.3334(1) & 1 & -0.0772(8) & 0.004(1) & - & - & - & - & - & - \\
 & 2048 & 13.4/21 & 0.97389477(7) & 1.3339(1) & 1 & -0.0776(6) & 0.005(1) & - & - & - & - & - & - \\ 
 & 1024 & 31.8/27 & 0.9738957(2) & 1.3333(1) & 1 & -0.0762(7) & - & - & - & - & - & - & - \\
 & 2048 & 22.4/22 & 0.9738954(2) & 1.3337(2) & 1 & -0.0762(6) & - & - & - & - & - & - & - \\
\hline
1.5 
 & 8192 & 16.5/20 & 0.950482391(7) & 1.3353(3) & 1 & -0.0544(2) & 0.0029(4) & -0.01(3) & - & 0.5 & - & - & - \\
 & 16384 & 11.5/15 & 0.950482462(9) & 1.3347(5) & 1 & -0.0545(2) & 0.0030(4) & 0.07(6) & - & 0.5 & - & - & - \\
 & 8192 & 11.5/19 & 0.950482382(6) & 1.3354(2) & 1 & -0.0555(4) & 0.0027(4) & -0.02(2) & - & 0.5 & - & 0.18(6) & - \\
 & 16384 & 9.7/14 & 0.950482456(8) & 1.3347(4) & 1 & -0.0554(6) & 0.0028(4) & 0.07(6) & - & 0.5 & - & 0.2(1) & - \\
 & 2048 & 14.1/28 & 0.950482439(6) & 1.3345(2) & 1 & -0.0553(3) & 0.0027(3) & 0.22(3) & -14.7(9) & 0.5 & 1 & 0.14(4) & - \\
 & 4096 & 11.4/23 & 0.950482529(7) & 1.3335(4) & 1 & -0.0553(3) & 0.0029(4) & 0.38(6) & -22(3) & 0.5 & 1 & 0.15(4) & - \\
 & 8192 & 10.2/18 & 0.950482525(9) & 1.3336(8) & 1 & -0.0555(4) & 0.0029(4) & 0.4(2) & -21(10) & 0.5 & 1 & 0.18(6) & - \\
 & 2048 & 20.8/29 & 0.950482447(6) & 1.3344(2) & 1 & -0.0544(2) & 0.0030(4) & 0.22(3) & -15(1) & 0.5 & 1 & - & - \\
 & 4096 & 17.3/24 & 0.950482542(7) & 1.3334(4) & 1 & -0.0543(2) & 0.0032(4) & 0.40(8) & -22(3) & 0.5 & 1 & - & - \\
 & 2048 & 28.3/30 & 0.950482585(5) & 4/3 & 1 & -0.0543(2) & 0.0035(4) & 0.351(8) & -18.6(4) & 0.5 & 1 & - & - \\
 & 4096 & 17.3/25 & 0.950482550(5) & 4/3 & 1 & -0.0543(2) & 0.0032(3) & 0.40(1) & -22(1) & 0.5 & 1 & - & - \\
 & 8192 & 14.7/20 & 0.950482549(5) & 4/3 & 1 & -0.0544(2) & 0.0031(4) & 0.41(4) & -23(3) & 0.5 & 1 & - & - \\
 & 2048 & 22.9/29 & 0.950482585(5) & 4/3 & 1 & -0.0551(4) & 0.0033(4) & 0.353(7) & -18.6(4) & 0.5 & 1 & 0.12(5) & - \\
 & 4096 & 11.5/24 & 0.950482548(4) & 4/3 & 1 & -0.0553(3) & 0.0029(3) & 0.41(1) & -22.6(8) & 0.5 & 1 & 0.15(4) & - \\
\hline\hline
\end{tabular*}
\end{table*}

\begin{table*}
\caption{Results of nonlinear fits of the mean walk length $N$ to Eq.~\eqref{eqn} for $\sigma=2.5$, $2.0$, $1.75$, and $1.5$.}
\label{n_large}
\begin{tabular*}{\textwidth}{@{\extracolsep{\fill}}llllllllll}
\hline\hline
$\sigma$ & $L_{\min}$ & $\chi^2$/DF & $1/\nu$ & $a$ & $b_1$ & $b_2$ & $y_1$ & $y_2$ & $c$ \\
\hline
 2.5
 & 256 & 4.4/4 & 0.9994(3) & 0.453(1) & - & - & - & - & -0.07(9)\\
 & 384 & 4.3/3 & 0.9993(5) & 0.454(2) & - & - & - & - & -0.1(2)\\
 & 512 & 4.1/2 & 0.9994(8) & 0.453(3) & - & - & - & - & -0.0(4)\\
 & 64 & 17.8/6 & 1.0000(6) & 0.451(2) & 0.1(1) & - & 0.9(4) & - & -\\
 & 128 & 10.5/5 & 0.9999(3) & 0.4511(8) & 0.14(4) & - & 1 & - & - \\
 & 256 & 4.4/4 & 0.9994(3) & 0.453(1) & -0.07(9) & - & 1 & - & - \\
 & 384 & 4.3/3 & 0.9993(5) & 0.454(2) & -0.1(2) & - & 1 & - & - \\
 & 16 & 20.5/9 & 0.9999(1) & 0.4512(4) & 0.163(9) & -1.41(9) & 1 & 2 & -\\
 & 32 & 18.8/8 & 1.0000(2) & 0.4508(6) & 0.18(2) & -1.8(4) & 1 & 2 & -\\
 \hline
2.0 
  & 512 & 1.3/4 & 0.99990(9) & 0.3959(3) & - & - & - & - & 1.53(6)\\
  & 1024 & 1.0/3 & 1.0000(1) & 0.3955(5) & - & - & - & - & 1.6(1)\\
  & 96 & 6.0/7 & 1.0006(2) & 0.3930(8) & 0.41(3) & - & 0.74(2) & - & -\\
  & 128 & 2.6/6 & 1.0003(2) & 0.3940(6) & 0.49(4) & - & 0.78(2) & - & -\\
  & 256 & 2.5/5 & 1.0004(3) & 0.394(1) & 0.5(1) & - & 0.77(6) & - & -\\
  & 384 & 1.5/4 & 1.0001(3) & 0.395(1) & 0.8(4) & - & 0.88(9) & - & -\\
  & 512 & 1.2/3 & 1.0000(3) & 0.395(1) & 1.2(8) & - & 0.9(1) & - & -\\
  & 384 & 2.1/5 & 0.99982(8) & 0.3961(3) & 1.46(4) & - & 1 & - & -\\
  & 512 & 1.3/4 & 0.99990(9) & 0.3959(3) & 1.53(6) & - & 1 & - & -\\
  & 1024 & 1.0/3 & 1.0000(1) & 0.3955(5) & 1.6(1) & - & 1 & - & -\\
   & 128 & 4.4/6 & 0.9998(1) & 0.3962(4) & 1.51(6) & -33(5) & 1 & 2 & -\\
  & 256 & 1.1/5 & 1.00004(8) & 0.3953(3) & 1.77(8) & -82(13) & 1 & 2 & -\\
 \hline
 1.75

  & 32 & 10.9/9 & 1.0004(3) & 0.3373(9) & 0.88(3) & -2.3(2) & 0.691(9) & 2$y_1$ & - \\
  & 48 & 6.6/8 & 1.0000(3) & 0.3386(9) & 0.96(4) & -3.0(4) & 0.71(1) & 2$y_1$ & - \\
  & 64 & 5.4/7 & 0.9997(3) & 0.340(1) & 1.04(8) & -3.8(8) & 0.73(2) & 2$y_1$ & - \\
 
  & 48 & 12.6/8 & 1.0007(4) & 0.336(1) & 0.66(4) & -10(1) & 0.64(2) & 2 & -\\
  & 64 & 7.6/7 & 1.0002(4) & 0.338(1) & 0.75(6) & -14(2) & 0.67(2) & 2 & -\\
  & 96 & 6.1/6 & 0.9998(5) & 0.339(2) & 0.9(1) & -22(7) & 0.70(3) & 2 & -\\
   & 64 & 13.7/8 & 1.0005(2) & 0.3370(6) & 1.62(3) & - & 0.75 & - & -1.88(7)\\
  & 96 & 8.5/7 & 1.0002(2) & 0.3380(7) & 1.54(5) & - & 0.75 & - & -1.6(1)\\
  & 128 & 7.8/6 & 1.0001(3) & 0.3383(9) & 1.50(7) & - & 0.75 & - & -1.5(2)\\
   & 96 & 8.6/7 & 1.0002(2) & 0.3380(7) & 1.54(5) & -1.6(1) & 0.75 & 1 & -\\
  & 128 & 7.8/6 & 1.0001(3) & 0.3383(9) & 1.50(7) & -1.5(2) & 0.75 & 1 & -\\
  & 256 & 1.6/5 & 0.9994(2) & 0.3408(7) & 1.16(9) & -0.4(3) & 0.75 & 1 & -\\
   & 32 & 9.2/9 & 1.0001(2) & 0.3382(6) & 1.46(5) & -1.5(3) & 0.75 & 1.5 & -1.3(2)\\
  & 48 & 6.2/8 & 0.9999(2) & 0.3390(7) & 1.35(7) & -2.5(5) & 0.75 & 1.5 & -0.9(3)\\
  & 64 & 5.3/7 & 0.9997(3) & 0.3397(9) & 1.3(1) & -3(1) & 0.75 & 1.5 & -0.5(4)\\
 \hline
1.5 
  & 64 & 8.0/7 & 1.0009(6) & 0.234(2) & 1.12(3) & - & 0.518(6) & - & -1.67(7)\\
  & 96 & 7.8/6 & 1.0006(8) & 0.235(2) & 1.14(6) & - & 0.52(1) & - & -1.7(2)\\
  & 128 & 7.0/5 & 1.000(1) & 0.236(4) & 1.2(1) & - & 0.53(2) & - & -1.9(3)\\
  & 256 & 7.0/4 & 1.000(2) & 0.236(7) & 1.2(3) & - & 0.53(5) & - & -2(1)\\

   & 96 & 8.0/6 & 1.0009(7) & 0.234(2) & 1.08(3) & -1.7(2) & 0.514(7) & 2$y_1$ & - \\
  & 128 & 7.1/5 & 1.000(1) & 0.236(3) & 1.13(6) & -2.0(4) & 0.52(1) & 2$y_1$ & - \\
  & 256 & 7.0/4 & 1.000(2) & 0.237(6) & 1.2(2) & -2(1) & 0.53(3) & 2$y_1$ & - \\
 
  & 64 & 8.0/7 & 1.0009(6) & 0.234(2) & 1.12(3) & -1.67(7) & 0.518(6) & 1 & -\\
  & 96 & 7.8/6 & 1.0006(8) & 0.235(2) & 1.14(6) & -1.7(2) & 0.52(1) & 1 & -\\
  & 128 & 7.0/5 & 1.000(1) & 0.236(4) & 1.2(1) & -1.9(3) & 0.53(2) & 1 & -\\
 
   & 1024 & 2.8/3 & 1.0002(3) & 0.2356(9) & 0.922(7) & - & 0.5 & - & -\\
 
    & 128 & 11.1/6 & 1.0021(3) & 0.2304(8) & 1.019(6) & - & 0.5 & - & -1.41(4)\\
  & 256 & 8.5/5 & 1.0017(5) & 0.231(1) & 1.01(1) & - & 0.5 & - & -1.3(1)\\
  & 384 & 8.2/4 & 1.0015(7) & 0.232(2) & 1.00(2) & - & 0.5 & - & -1.2(2)\\
 
   & 128 & 11.2/6 & 1.0022(3) & 0.2304(8) & 1.020(6) & -1.41(4) & 0.5 & 1 & -\\
  & 256 & 8.5/5 & 1.0017(5) & 0.231(1) & 1.01(1) & -1.3(1) & 0.5 & 1 & -\\
  & 384 & 8.2/4 & 1.0015(7) & 0.232(2) & 1.00(2) & -1.2(2) & 0.5 & 1 & -\\
  & 512 & 8.2/3 & 1.001(1) & 0.232(3) & 0.99(4) & -1.2(5) & 0.5 & 1 & -\\
\hline\hline
\end{tabular*}
\end{table*}

\begin{table*}
\caption{Results of nonlinear fits of the empty-walk probability $D_0$ to Eq.~(\ref{d0eq}) for $\sigma=2.5$, $2.0$, $1.75$, and $1.5$. }
\label{d_large}
\begin{tabular*}{\textwidth}{@{\extracolsep{\fill}}lllllllll}
\hline\hline
$\sigma$ & $L_{\min}$ & $\chi^2$/DF & $\eta$ & $a$ & $b_1$ & $b_2$ & $y_1$ & $y_2$ \\
\hline
2.5 

  & 512 & 3.0/3 & 0.9995(6) & 0.457(2) & - & - & - & -\\
  & 1024 & 2.5/2 & 1.000(1) & 0.455(4) & - & - & - & -\\
   & 48 & 9.5/7 & 0.9995(8) & 0.457(3) & 0.5(3) & - & 1.1(2) & -\\
  & 64 & 9.1/6 & 1.000(1) & 0.455(4) & 0.4(3) & - & 0.9(3) & -\\
   & 128 & 7.0/5 & 0.9997(7) & 0.456(2) & 0.5(1) & - & 1 & -\\
  & 256 & 2.6/4 & 1.0013(7) & 0.450(3) & 1.1(2) & - & 1 & -\\
  & 384 & 2.6/3 & 1.001(1) & 0.450(5) & 1.1(6) & - & 1 & -\\
   & 32 & 9.3/7 & 1.000(2) & 0.454(6) & 0.1(3) & 0.6(7) & 0.6(5) & 2$y_1$\\
  & 48 & 9.1/6 & 1.000(2) & 0.454(7) & 0.1(7) & 1(3) & 1(1) & 2$y_1$\\
 \hline
 2.0
   & 4096 & 0.3/2 & 0.9988(2) & 0.4124(8) & - & - & - & -\\
  & 8192 & 0.2/1 & 0.9991(5) & 0.411(2) & - & - & - & -\\
  & 128 & 2.9/6 & 0.9996(5) & 0.409(2) & 0.4(1) & - & 0.76(8) & -\\
  & 256 & 1.8/5 & 0.9987(4) & 0.412(2) & 2(2) & - & 1.1(2) & -\\
  & 384 & 1.1/4 & 1.000(1) & 0.408(5) & 0.2(2) & - & 0.6(3) & -\\
  & 512 & 0.4/3 & 1.002(4) & 0.40(2) & 0.06(2) & - & 0.3(3) & -\\
   & 48 & 7.8/8 & 0.999(2) & 0.411(7) & 0.0(1) & 0.7(2) & 0.5(1) & 2$y_1$\\
  & 64 & 4.9/7 & 0.9993(9) & 0.410(4) & 0.1(2) & 1.0(7) & 0.5(2) & 2$y_1$\\
 \hline
 1.75
  & 256 & 1.8/5 & 1.003(2) & 0.350(6) & 0.39(9) & - & 0.53(7) & -\\
  & 384 & 0.3/4 & 1.0012(6) & 0.358(2) & 0.8(2) & - & 0.68(5) & -\\
  & 512 & 0.1/3 & 1.0007(4) & 0.360(2) & 1.1(2) & - & 0.75(4) & -\\
  & 1024 & 0.1/2 & 1.001(1) & 0.357(4) & 0.6(4) & - & 0.6(1) & -\\
  & 8 & 5.8/12 & 0.999(1) & 0.371(5) & -0.07(2) & 0.62(1) & 0.290(5) & 2$y_1$\\
  & 16 & 4.2/11 & 1.001(1) & 0.361(5) & -0.02(3) & 0.59(2) & 0.31(1) & 2$y_1$\\
  & 64 & 2.4/7 & 1.003(2) & 0.352(7) & 0.3(2) & 0.5(3) & 0.5(1) & 1\\
  & 96 & 2.4/6 & 1.003(3) & 0.35(1) & 0.3(3) & 0.5(7) & 0.5(2) & 1\\
  & 128 & 2.4/5 & 1.003(4) & 0.35(2) & 0.3(4) & 1(1) & 0.5(3) & 1\\
   & 128 & 10.0/7 & 0.9991(4) & 0.366(1) & 0.88(2) & - & 0.75 & -\\
  & 256 & 5.7/6 & 0.9997(4) & 0.364(1) & 0.95(3) & - & 0.75 & -\\
  & 384 & 0.5/5 & 1.0004(2) & 0.3613(5) & 1.05(2) & - & 0.75 & -\\
  & 32 & 6.7/10 & 0.9998(3) & 0.3634(8) & 1.13(3) & -0.73(5) & 0.75 & 1\\
  & 48 & 5.6/9 & 1.0000(3) & 0.363(1) & 1.18(4) & -0.84(9) & 0.75 & 1\\
  & 64 & 3.9/8 & 1.0003(3) & 0.362(1) & 1.25(5) & -1.0(1) & 0.75 & 1\\
  & 96 & 3.0/7 & 1.0007(4) & 0.360(1) & 1.4(1) & -1.3(2) & 0.75 & 1\\
   & 64 & 7.6/8 & 0.9994(3) & 0.365(1) & 0.92(2) & -12(2) & 0.75 & 2\\
  & 96 & 5.0/7 & 0.9998(4) & 0.363(1) & 0.97(3) & -22(6) & 0.75 & 2\\
  & 128 & 3.8/6 & 1.0002(4) & 0.362(1) & 1.02(5) & -36(12) & 0.75 & 2\\
 \hline
1.5
  & 128 & 5.5/6 & 1.002(1) & 0.272(4) & 0.68(1) & - & 0.444(8) & -\\
  & 256 & 2.8/5 & 0.999(1) & 0.281(5) & 0.74(3) & - & 0.47(1) & -\\
  & 384 & 2.6/4 & 0.998(2) & 0.284(7) & 0.78(8) & - & 0.49(3) & -\\
  & 512 & 2.5/3 & 0.998(3) & 0.29(1) & 0.8(1) & - & 0.49(5) & -\\
   & 128 & 2.9/5 & 0.996(2) & 0.291(7) & 1.1(2) & -1(1) & 0.54(4) & 2$y_1$\\
   & 32 & 4.5/9 & 1.0004(9) & 0.277(3) & 0.75(2) & -0.24(4) & 0.467(8) & 1\\
  & 48 & 4.5/8 & 1.000(1) & 0.278(4) & 0.76(4) & -0.26(9) & 0.47(1) & 1\\
  & 64 & 4.4/7 & 1.000(2) & 0.279(5) & 0.78(6) & -0.3(1) & 0.48(2) & 1\\
  & 96 & 4.1/6 & 0.999(2) & 0.282(7) & 0.8(1) & -0.5(3) & 0.49(3) & 1\\
   & 1024 & 2.4/3 & 0.997(1) & 0.288(3) & 0.81(3) & - & 0.5 & -\\
  & 2048 & 0.8/2 & 0.999(1) & 0.282(4) & 0.89(4) & - & 0.5 & -\\
   & 384 & 2.6/4 & 0.998(1) & 0.286(4) & 0.83(5) & -0.2(5) & 0.5 & 1\\
  & 512 & 2.6/3 & 0.997(2) & 0.287(6) & 0.82(9) & 0(1) & 0.5 & 1\\
  & 1024 & 0.4/2 & 1.002(2) & 0.271(6) & 1.2(1) & -5(2) & 0.5 & 1\\
\hline\hline
\end{tabular*}
\end{table*}

\begin{table*}[t]
\centering
\caption{Results of nonlinear fits of the Binder ratio $Q_N(z,L)$ to Eq.~\eqref{eq:fss_fitting_ansatz} for $\sigma=1.0$, $0.9$, $0.75$, and $0.6$.}
\label{tabqn}
\footnotesize
\setlength{\tabcolsep}{2pt}
\begin{tabularx}{\textwidth}{@{\extracolsep{\fill}}l l l l l l l l l l l l l }
\hline\hline
$\sigma$ & $L_{\rm min}$ & $\chi^2/{\rm DF}$ & \multicolumn{1}{l}{$z_c$} & \multicolumn{1}{l}{$Q_N^c$} & \multicolumn{1}{l}{$y_t (1/\nu)$} & \multicolumn{1}{l}{$a_1$} & \multicolumn{1}{l}{$a_2$}  & \multicolumn{1}{l}{$b_1$}  & \multicolumn{1}{l}{$b_2$}& \multicolumn{1}{l}{$y_1$}  & \multicolumn{1}{l}{$y_2$} & \multicolumn{1}{l}{$c_1$} \\
\hline
1.0 & 128 & 56.5/54 & 0.8108897(5) & 1.29049(6) & 0.660(2) & -0.136(2) & 0.007(2) & 0.274(7) & - & 0.620(5) & -& - \\
& 256 & 50.3/48 & 0.8108902(5) & 1.29039(8) & 0.660(2) & -0.137(2) & 0.007(2) & 0.26(2) & - & 0.61(2) & -& - \\
& 512 & 39.8/42 & 0.8108896(7) & 1.2906(2) & 0.661(2) & -0.135(2) & 0.007(2) & 0.29(4) & - & 0.63(3) & -& - \\
& 1024 & 36.5/36 & 0.8108897(9) & 1.2905(2) & 0.661(2) & -0.135(2) & 0.006(2) & 0.26(8) & - & 0.61(5) & -& - \\

& 16 & 105.5/72 & 0.8108892(3) & 1.29056(2) & 0.658(2) & -0.139(2) & 0.008(2) & 0.2676(3) & 0.570(4) & 0.62 & 1.5& - \\
& 32 & 91.2/66 & 0.8108892(3) & 1.29056(2) & 0.660(2) & -0.138(2) & 0.008(2) & 0.2681(6) & 0.56(2) & 0.62 & 1.5& - \\
& 64 & 74.4/60 & 0.8108893(4) & 1.29055(3) & 0.660(2) & -0.137(2) & 0.007(2) & 0.269(2) & 0.49(4) & 0.62 & 1.5& - \\
& 128 & 56.3/54 & 0.8108897(4) & 1.29050(4) & 0.660(2) & -0.136(2) & 0.007(2) & 0.274(2) & 0.5(2) & 0.62 & 1.5& - \\
& 256 & 48.6/48 & 0.8108901(4) & 1.29043(5) & 0.660(2) & -0.136(2) & 0.007(2) & 0.281(4) & -0.9(5) & 0.62 & 1.5& - \\

& 8 & 121.1/77 & 0.8108887(3) & 1.29063(2) & 0.666(2) & -0.128(2) & 0.006(2) & 0.2654(2) & 0.598(2) & 0.62 & 1.5& -0.27(4) \\
& 16 & 66.9/71 & 0.8108893(3) & 1.29057(2) & 0.665(2) & -0.128(3) & 0.006(2) & 0.2669(4) & 0.579(4) & 0.62 & 1.5& -0.27(5) \\
& 32 & 66.8/65 & 0.8108893(3) & 1.29057(2) & 0.665(2) & -0.128(3) & 0.006(2) & 0.2669(6) & 0.58(2) & 0.62 & 1.5& -0.27(6) \\
& 64 & 64.3/59 & 0.8108893(4) & 1.29057(3) & 0.664(2) & -0.130(3) & 0.006(2) & 0.268(2) & 0.55(5) & 0.62 & 1.5& -0.22(7) \\
& 128 & 53.4/53 & 0.8108896(4) & 1.29051(4) & 0.663(3) & -0.131(4) & 0.006(2) & 0.272(3) & 0.2(2) & 0.62 & 1.5& -0.2(1) \\

& 8 & 121.1/78 & 0.8108887(3) & 1.29063(2) & 0.6662 & -0.1270(3) & 0.0058(8) & 0.2654(2) & 0.598(2) & 0.62 & 1.5& -0.28(3) \\
& 16 & 67.3/72 & 0.8108893(3) & 1.29057(2) & 0.6662 & -0.1271(3) & 0.0051(8) & 0.2669(3) & 0.580(4) & 0.62 & 1.5& -0.29(4) \\
& 32 & 67.2/66 & 0.8108893(3) & 1.29057(2) & 0.6662 & -0.1271(3) & 0.0051(8) & 0.2668(6) & 0.59(2) & 0.62 & 1.5& -0.29(4) \\
& 64 & 65.6/60 & 0.8108892(4) & 1.29057(3) & 0.6662 & -0.1271(3) & 0.0050(8) & 0.267(2) & 0.57(5) & 0.62 & 1.5& -0.27(5) \\
& 128 & 55.4/54 & 0.8108896(4) & 1.29053(3) & 0.6662 & -0.1272(3) & 0.0048(9) & 0.271(2) & 0.3(2) & 0.62 & 1.5& -0.26(6) \\
\hline
0.9 & 32 & 92.3/66 & 0.7667048(3) & 1.31722(2) & 0.6293(9) & -0.160(2) & 0.005(2) & 0.432(2) & - & 0.725(2) & 1.5& - \\
& 64 & 78.6/60 & 0.7667051(4) & 1.31718(3) & 0.6296(9) & -0.160(2) & 0.004(2) & 0.424(4) & - & 0.721(3) & 1.5& - \\
& 128 & 59.7/54 & 0.7667048(4) & 1.31721(4) & 0.631(2) & -0.159(2) & 0.004(2) & 0.43(2) & - & 0.723(5) & 1.5& - \\
& 256 & 50.7/48 & 0.7667046(5) & 1.31724(5) & 0.631(2) & -0.158(2) & 0.004(2) & 0.44(3) & - & 0.73(1) & 1.5& - \\
& 512 & 49.5/42 & 0.7667044(6) & 1.31727(8) & 0.631(2) & -0.157(2) & 0.005(2) & 0.47(6) & - & 0.74(3) & 1.5& - \\

& 32 & 91.6/66 & 0.7667050(3) & 1.31719(2) & 0.6293(9) & -0.160(2) & 0.004(2) & 0.4220(8) & 0.05(2) & 0.72 & 1.5& - \\
& 64 & 78.4/60 & 0.7667052(3) & 1.31717(2) & 0.6297(9) & -0.160(2) & 0.004(2) & 0.425(2) & 0.04(1) & 0.72 & 1.5& - \\
& 128 & 60.2/54 & 0.7667050(4) & 1.31719(3) & 0.631(2) & -0.158(2) & 0.004(2) & 0.421(3) & 0.04(2) & 0.72 & 1.5& - \\
& 256 & 50.8/48 & 0.7667048(4) & 1.31721(4) & 0.631(2) & -0.158(2) & 0.004(2) & 0.416(7) & 0.5(3) & 0.72 & 1.5& - \\

& 32 & 66.2/65 & 0.7667050(3) & 1.31719(2) & 0.634(2) & -0.152(3) & 0.004(2) & 0.4211(8) & 0.07(2) & 0.72 & 1.5& -0.43(9) \\
& 64 & 59.7/59 & 0.7667051(3) & 1.31718(2) & 0.634(2) & -0.152(3) & 0.004(2) & 0.423(2) & 0.04(2) & 0.72 & 1.5& -0.5(2) \\
& 128 & 54.9/53 & 0.7667049(4) & 1.31720(3) & 0.634(2) & -0.153(3) & 0.004(2) & 0.419(4) & 0.16(7) & 0.72 & 1.5& -0.4(2) \\
& 256 & 49.8/47 & 0.7667047(4) & 1.31722(4) & 0.633(2) & -0.155(3) & 0.004(2) & 0.415(7) & 0.4(2) & 0.72 & 1.5& -0.2(3) \\

& 32 & 66.3/66 & 0.7667050(3) & 1.31719(2) & 0.6332 & -0.1531(2) & 0.004(2) & 0.4212(8) & 0.07(2) & 0.72 & 1.5& -0.42(7) \\
& 64 & 60.0/60 & 0.7667051(3) & 1.31718(2) & 0.6332 & -0.1530(2) & 0.004(2) & 0.423(2) & 0.04(4) & 0.72 & 1.5& -0.44(8) \\
& 128 & 54.9/54 & 0.7667049(4) & 1.31720(3) & 0.6332 & -0.1531(2) & 0.004(2) & 0.419(4) & 0.16(7) & 0.72 & 1.5& -0.4(2) \\
& 256 & 50.3/48 & 0.7667047(4) & 1.31722(4) & 0.6332 & -0.1531(2) & 0.004(2) & 0.414(7) & 0.5(3) & 0.72 & 1.5& -0.4(2) \\

\hline
0.75 & 64 & 83.9/78 & 0.6995730(4) & 1.3742(2) & 0.5903(8) & -0.206(2) & - & -0.083(2) & 0.518(5) & 0.318(3) & $2y_1$ & - \\
& 128 & 75.8/71 & 0.6995729(5) & 1.3743(2) & 0.5906(8) & -0.205(2) & - & -0.083(2) & 0.51(2) & 0.316(4) & $2y_1$ & - \\
& 256 & 67.5/64 & 0.6995723(7) & 1.3746(4) & 0.5910(8) & -0.204(2) & - & -0.083(2) & 0.46(3) & 0.303(8) & $2y_1$ & - \\
& 512 & 65.1/57 & 0.699572(2) & 1.3750(6) & 0.5912(9) & -0.204(2) & - & -0.083(2) & 0.42(6) & 0.30(2) & $2y_1$ & - \\
& 1024 & 48.3/50 & 0.699575(1) & 1.3730(5) & 0.5913(9) & -0.203(2) & - & -0.11(2) & 1.4(6) & 0.40(4) & $2y_1$ & - \\
& 2048 & 35.2/43 & 0.699573(2) & 1.375(3) & 0.592(1) & -0.202(3) & - & -0.07(2) & 0.3(3) & 0.26(9) & $2y_1$ & - \\

& 64 & 84.6/79 & 0.6995733(3) & 1.37409(4) & 0.5902(8) & -0.206(2) & - & -0.0816(5) & 0.522(2) & 0.32 & $2y_1$& - \\
& 128 & 76.9/72 & 0.6995733(3) & 1.37407(5) & 0.5906(8) & -0.205(2) & - & -0.0813(7) & 0.519(3) & 0.32 & $2y_1$& - \\
& 256 & 71.9/65 & 0.6995734(3) & 1.37404(7) & 0.5909(8) & -0.204(2) & - & -0.081(2) & 0.516(5) & 0.32 & $2y_1$& - \\
& 512 & 68.7/58 & 0.6995732(4) & 1.3742(1) & 0.5911(9) & -0.204(2) & - & -0.083(2) & 0.53(1) & 0.32 & $2y_1$& - \\
& 1024 & 54.1/51 & 0.6995723(5) & 1.3744(2) & 0.5914(9) & -0.203(2) & - & -0.088(4) & 0.56(3) & 0.32 & $2y_1$& - \\
& 2048 & 35.8/44 & 0.6995734(7) & 1.3740(3) & 0.592(1) & -0.203(3) & - & -0.078(7) & 0.49(5) & 0.32 & $2y_1$& - \\

& 64 & 84.0/79 & 0.6995730(4) & 1.3742(2) & 0.5900 & -0.2058(2) & - & -0.083(2) & 0.518(5) & 0.318(3) & $2y_1$& - \\
& 128 &76.4/72 & 0.6995729(5) & 1.3743(2) & 0.5900 & -0.2058(2) & - & -0.083(2) & 0.51(2) & 0.316(4) & $2y_1$& - \\
& 256 & 69.2/65 & 0.6995723(7) & 1.3746(4) & 0.5900 & -0.2058(2) & - & -0.083(2) & 0.46(3) & 0.304(8) & $2y_1$& - \\
& 512 & 67.3/58 & 0.699572(1) & 1.3750(6) & 0.5900 & -0.2058(2) & - & -0.083(2) & 0.42(6) & 0.30(2) & $2y_1$& - \\
& 1024 & 50.5/51 & 0.699575(1) & 1.3730(5) & 0.5900 & -0.2058(2) & - & -0.11(2) & 1.4(6) & 0.41(4) & $2y_1$& - \\
& 2048 & 37.8/44 & 0.699573(2) & 1.375(3) & 0.5900 & -0.2058(2) & - & -0.07(2) & 0.3(3) & 0.27(9) & $2y_1$& - \\
\hline
0.6 & 128 & 113/92 & 0.6361373(5) & 1.4462(3) & 0.5459(5) & -0.272(2) & -0.016(3) & -0.2060(7) & 0.572(5) & 0.256(2) & $2y_1$ & - \\
& 256 & 79.3/83 & 0.6361355(7) & 1.4475(4) & 0.5461(5) & -0.271(2) & -0.015(3) & -0.2055(8) & 0.53(1) & 0.245(3) & $2y_1$& - \\
& 512 & 71.1/74 & 0.6361351(9) & 1.4479(7) & 0.5464(5) & -0.270(2) & -0.015(3) & -0.204(2) & 0.51(3) & 0.241(5) & $2y_1$& - \\
& 1024 & 62.1/65 & 0.636134(2) & 1.450(2) & 0.5467(5) & -0.270(2) & -0.014(3) & -0.196(4) & 0.43(4) & 0.23(1) & $2y_1$& - \\

& 128 & 128.3/93 & 0.6361356(3) & 1.44708(5) & 0.5459(5) & -0.272(2) & -0.015(3) & -0.2078(5) & 0.555(2) & 0.25 & $2y_1$& - \\
& 256 & 82.2/84 & 0.6361364(3) & 1.44686(7) & 0.5461(5) & -0.271(2) & -0.016(3) & -0.2052(8) & 0.548(3) & 0.25 & $2y_1$& - \\
& 512 & 74.1/75 & 0.6361364(4) & 1.4469(1) & 0.5463(5) & -0.271(2) & -0.015(3) & -0.206(2) & 0.548(4) & 0.25 & $2y_1$& - \\
& 1024 & 69.9/66 & 0.6361364(5) & 1.4469(2) & 0.5466(5) & -0.270(2) & -0.015(3) & -0.205(3) & 0.548(8) & 0.25 & $2y_1$& - \\

& 128 & 113.1/93 & 0.6361373(5) & 1.4462(3) & 0.5459 & -0.2715(2) & -0.016(2) & -0.2060(7) & 0.572(5) & 0.256(2) &$2y_1$& - \\
& 256 & 79.5/84 & 0.6361355(7) & 1.4475(4) & 0.5459 & -0.2715(2) & -0.016(2) & -0.2056(8) & 0.54(1) & 0.246(3) & $2y_1$& - \\
& 512 & 72.2/75 & 0.6361351(9) & 1.4479(7) & 0.5459 & -0.2715(2) & -0.016(2) & -0.205(2) & 0.51(3) & 0.242(5) & $2y_1$& - \\
& 1024 & 64.7/66 & 0.636134(2) & 1.450(2) & 0.5459 & -0.2714(2) & -0.016(2) & -0.196(4) & 0.50(2) & 0.23(1) & $2y_1$& - \\

\hline\hline
\end{tabularx}
\end{table*}

\begin{table*}[t]
\caption{Results of nonlinear fits of the mean walk length $N$ to Eq.~\eqref{eqn} for $\sigma=1.0$, $0.9$, $0.75$, and $0.6$. }
\label{tabn}
\begin{tabular*}{\textwidth}{@{\extracolsep{\fill}}llllllllll}
\hline\hline
$\sigma$ & $L_{\min}$ & $\chi^2$/DF & $y_t (1/\nu)$ & $a$ & $b_1$ & $b_2$ & $y_1$ & $y_2$ & $c$ \\
\hline
1.0&4096 & 4.3/6 & 0.6660(2) & 0.989(2) & -1.16(6)& -& 0.6&-& - \\
&8192 & 4.3/5 & 0.6660(2) & 0.988(2) & -1.2(2)& -& 0.6&-& - \\
&16384 & 3.6/4 & 0.6662(4) & 0.985(4) & -0.9(3)& -& 0.6&-& - \\
&256 & 10.9/9 & 0.6660(8) & 0.9873(9) & -1.9(2)& -& 0.6&-& 1.2(3) \\
&512 & 6.1/8 & 0.6659(2) & 0.989(2) & -2.4(3)& -& 0.6&-& 1.9(4) \\
&1024 & 5.5/7 & 0.6657(2) & 0.991(3) & -2.8(6)& -& 0.6&-& 2.5(9) \\
&64 & 9.4/10 & 0.66588(8) & 0.9884(9) & -2.4(2)& -1.5(3)& 0.6&$2y_1$& 1.9(3) \\
&128 & 7.1/9 & 0.6658(2) & 0.990(2) & -2.8(4)& -2.4(7)& 0.6&$2y_1$& 2.6(6) \\
&256 & 6.0/8 & 0.6657(2) & 0.992(3) & -3.5(8)& -4(2)& 0.6&$2y_1$& 4(2) \\
\hline
0.9&4096 & 8.3/6 & 0.6335(2) & 1.016(2) & -1.22(8)& -& 0.6&-& - \\
&8192 & 6.3/5 & 0.6332(3) & 1.019(3) & -1.4(2)& -& 0.6&-& - \\
&16384 & 3.4/4 & 0.6328(4) & 1.025(5) & -1.9(4)& -& 0.6&-& - \\
&256 & 11.3/9 & 0.6335(2) & 1.016(2) & -2.7(4)& -& 0.6&-& 1.9(4) \\
&512 & 6.7/8 & 0.6333(2) & 1.019(2) & -4.0(7)& -& 0.6&-& 3.4(8) \\
&1024 & 6.6/7 & 0.6332(3) & 1.019(3) & -4(2)& -& 0.6&-& 3(2) \\
&2048 & 5.9/6 & 0.6330(4) & 1.022(5) & -6(3)& -& 0.6&-& 6(4) \\
&64 & 11.8/10 & 0.6335(2) & 1.017(2) & -3.4(5)& -1.1(3)& 0.6&$2y_1$& 2.7(5) \\
&128 & 7.7/9 & 0.6332(2) & 1.020(2) & -4.9(9)& -2.5(8)& 0.6&$2y_1$& 5(2) \\
&256 & 6.6/8 & 0.6330(3) & 1.022(3) & -7(2)& -5(3)& 0.6&$2y_1$& 7(3) \\
\hline
0.75&4096 & 3.2/6 & 0.59012(6) & 1.0393(7) & -1.20(3)& -& 0.56&-& - \\
&8192 & 2.3/5 & 0.59006(9) & 1.040(2) & -1.23(5)& -& 0.56&-& - \\
&16384 & 2.2/4 & 0.5901(2) & 1.040(2) & -1.20(9)& -& 0.56&-& - \\
&1024 & 4.2/7 & 0.5900(2) & 1.043(2) & -4.1(6)& -& 0.56&-& 3.5(6) \\
&2048 & 4.1/6 & 0.5900(2) & 1.043(2) & -5(1)& -& 0.56&-& 3.7(8) \\
&4096 & 2.8/5 & 0.5901(6) & 1.041(3) & -3(2)& -& 0.56&-& 1(1) \\
&64 & 15/10 & 0.59008(6) & 1.0388(8) & -2.4(3)& -0.6(2)& 0.56&$2y_1$& 1.5(3) \\
&128 & 10.1/9 & 0.58995(9) & 1.041(2) & -3.1(5)& -1.2(4)& 0.56&$2y_1$& 2.4(5) \\
&256 & 6.6/8 & 0.5899(2) & 1.043(2) & -4.4(8)& -3(1)& 0.56&$2y_1$& 4(1) \\
&512 & 4.2/7 & 0.5899(2) & 1.046(3) & -7(2)& -7(3)& 0.56&$2y_1$& 6(2) \\
\hline
0.6&2048 & 7.9/7 & 0.54602(7) & 1.0731(8) & -1.13(2)& -& 0.5&-& - \\
&4096 & 7.1/6 & 0.5460(2) & 1.074(2) & -1.15(3)& -& 0.5&-& - \\
&8192 & 6.3/5 & 0.5460(2) & 1.073(2) & -1.11(5)& -& 0.5&-& - \\
&16384 & 3.3/4 & 0.5460(2) & 1.069(3) & -0.99(9)& -& 0.5&-& - \\
&128 & 11.5/10 & 0.54585(6) & 1.0755(8) & -1.69(7)& -& 0.5&-& 0.73(8) \\
&256 & 8.9/9 & 0.54578(9) & 1.077(2) & -1.8(2)& -& 0.5&-& 0.9(2) \\
&512 & 8.1/8 & 0.5459(2) & 1.076(2) & -1.7(3)& -& 0.5&-& 0.7(3) \\
&16 & 13.4/12 & 0.54580(6) & 1.0763(7) & -1.91(7)& -0.25(3)& 0.5&$2y_1$& 1.01(8) \\
&32 & 10.8/11 & 0.54578(8) & 1.077(2) & -2.0(2)& -0.29(5)& 0.5&$2y_1$& 1.2(2) \\
&64 & 9.9/10 & 0.5458(2) & 1.078(2) & -2.1(2)& -0.4(2)& 0.5&$2y_1$& 1.3(3) \\
\hline\hline
\end{tabular*}
\end{table*}

\begin{table*}[t]
\caption{Results of nonlinear fits of the empty-walk probability $D_0$ to Eq.~(\ref{d0eq}) for $\sigma=1.0$, $0.9$, $0.75$, and $0.6$.  }
\label{tabd}
\begin{tabular*}{\textwidth}{@{\extracolsep{\fill}}lllllllll}
\hline\hline
$\sigma$ & $L_{\min}$ & $\chi^2$/DF & $\eta$  & $a$   & $b_1$  & $b_2$  & $y_1$ & $y_2$   \\ \hline
 1.0&32 & 12.4/12 & 0.9996(2) & 0.7393(6) & 0.136(2)&  -& 0.45&- \\
 &64 & 10.3/11 & 0.9995(2) & 0.7401(8) & 0.132(3)&  -& 0.45&- \\
 &128 & 9.8/10 & 0.9994(2) & 0.741(2) & 0.129(6)&  -& 0.45&- \\
 &256 & 8.8/9 & 0.9993(3) & 0.741(2) & 0.13(2)&  -& 0.45&- \\
 &512 & 6.4/8 & 0.9997(4) & 0.738(3) & 0.16(3)&  -& 0.45&- \\
 &8 & 17.8/13 & 0.9989(2) & 0.7447(7) & 0.099(3)&  0.100(5)& 0.45&0.9 \\
 &16 & 15.2/12 & 0.9990(2) & 0.744(2) & 0.106(6)&  0.09(2)& 0.45&0.9 \\
 &32 & 10.2/11 & 0.9993(2) & 0.742(2) & 0.13(2)&  0.04(3)& 0.45&0.9 \\
\hline
 0.9&128 & 9.9/10 & 1.1004(2) & 0.641(1) & 0.216(4)&  -& 0.41&- \\
 &256 & 8.6/9 & 1.1002(2) & 0.642(2) & 0.211(7)&  -& 0.41&- \\
 &512 & 8.2/8 & 1.1001(3) & 0.643(2) & 0.21(2)&  -& 0.41&- \\
 &1024 & 7.1/7 & 1.1003(4) & 0.641(3) & 0.22(2)&  -& 0.41&- \\
 &32 & 12.1/11 & 1.1002(2) & 0.643(2) & 0.201(8)&  0.07(2)& 0.41&0.8 \\
 &64 & 8.8/10 & 1.0998(3) & 0.644(2) & 0.19(2)&  0.13(4)& 0.41&0.8 \\
 &128 &8.5/9 & 1.1000(4) & 0.644(3) & 0.19(3)&  0.09(9)& 0.41&0.8 \\
\hline
 0.75&4096 & 6.5/5 & 1.2536(6) & 0.472(4) & 0.203(8)&  -& 0.2&- \\
 &8192 & 4.2/4 & 1.2526(9) & 0.479(6) & 0.19(2)&  -& 0.2&- \\
 &16384 & 1.8/3 & 1.251(2) & 0.50(1) & 0.15(3)&  -& 0.2&- \\
 &32768 & 1.2/2 & 1.250(3) & 0.50(2) & 0.13(5)&  -& 0.2&- \\
 &32 & 7.5/11 & 1.2500(3) & 0.502(2) & 0.103(4)&  0.293(4)& 0.2&0.5 \\
 &64 & 7.2/10 & 1.2499(3) & 0.502(3) & 0.101(5)&  0.295(6)& 0.2&0.5 \\
 &128 & 5.9/9 & 1.2496(4) & 0.505(3) & 0.095(8)&  0.31(2)& 0.2&0.5 \\
 &256 & 5.1/8 & 1.2500(6) & 0.502(5) & 0.11(2)&  0.30(2)& 0.2&0.5 \\
 &512 & 4.8/7 & 1.2497(9) & 0.504(7) & 0.10(2)&  0.30(4)& 0.2&0.5 \\
 &1024 &4.2/6 & 1.249(2) & 0.51(2) & 0.10(4)&  0.35(8)& 0.2&0.5 \\
\hline
 0.6&16384 & 4.7/5 & 1.399(6) & 0.31(4) & 0.26(5)&  -& 0.14&- \\
 &64 & 7.0/9 & 1.4004(8) & 0.297(5) & 0.258(6)&  0.345(5)& 0.14&0.5 \\
 &128 & 6.8/8 & 1.400(1) & 0.299(6) & 0.255(9)&  0.348(9)& 0.14&0.5 \\
 &256 & 6.0/7 & 1.400(2) & 0.305(9) & 0.25(2)&  0.36(2)& 0.14&0.5 \\
 &512 & 3.2/6 & 1.397(3) & 0.32(2) & 0.22(2)&  0.41(4)& 0.14&0.5 \\
 &1024 & 2.8/5 & 1.396(4) & 0.33(2) & 0.21(4)&  0.44(7)& 0.14&0.5 \\
\hline\hline
\end{tabular*}
\end{table*}

\end{document}